\documentclass[12pt,a4paper]{article}
\usepackage{amssymb,booktabs,subfig,microtype,setspace,bm,paralist, dsfont, longtable} 
\usepackage[table,xcdraw]{xcolor}
\usepackage[top=1in, bottom=1in, left=1in, right=1in]{geometry}
\usepackage{natbib}
\usepackage{url}
\usepackage[pdftex,colorlinks=true,hypertexnames=false]{hyperref}
\definecolor{darkblue}{rgb}{0,0,.6}
\hypersetup{citecolor=darkblue,linkcolor=darkblue,urlcolor=darkblue}
\usepackage{amsmath}
\usepackage{amsfonts}
\usepackage{epsfig}
\usepackage{graphics}
\usepackage{palatino,eulervm}
\usepackage{graphicx}
\usepackage{lscape}
\usepackage{rotating}
\usepackage{multirow}
\usepackage[linewidth=1pt]{mdframed}
\usepackage{orcidlink}
\allowdisplaybreaks[4]

\usepackage{booktabs}
\usepackage{makecell}  
\usepackage{graphicx}  

\providecommand{\U}[1]{\protect\rule{.1in}{.1in}}

\graphicspath{{plots/}}

\usepackage{amsthm,thmtools}
\usepackage{mathrsfs}

\makeatletter
\def\th@newremark{\th@remark\thm@headfont{\bfseries}}
\makeatletter
\theoremstyle{newremark}

\declaretheoremstyle[
  spaceabove=6pt, spacebelow=6pt,
  headfont=\bfseries,
  notefont=\mdseries, notebraces={(}{)},
bodyfont=\normalfont,
  postheadspace=0.5em]{mystyle}

\newsavebox\CBox

\begin{document}

\title{\Large Extending finite mixture models with skew-normal distributions and hidden Markov models for time series}
\author{\normalsize Andrea Nigri \orcidlink{0000-0002-2707-3678} \qquad Marco Forti \orcidlink{0000-0003-2670-4869}\\
\normalsize Department of Economics, Management and Territory \\ 
\normalsize University of Foggia, Italy \\
\\ \normalsize Han Lin Shang \orcidlink{0000-0003-1769-6430}\\
\normalsize Department of Actuarial Studies and Business Analytics \\
\normalsize Macquarie University
}

\date{}

\maketitle

\centerline{\bf Abstract}
\{Highlight: \textit{We develop a novel Bayesian framework that integrates skew-normal mixtures into Hidden Markov models, and demonstrate its performance in regime-switching detection through extensive simulations and an empirical application to study gender gap in mortality data.}\}\\

We introduce an extension of finite mixture models by incorporating skew-normal distributions within a Hidden Markov Model framework. By assuming a constant transition probability matrix and allowing emission distributions to vary according to hidden states, the proposed model effectively captures dynamic dependencies between variables. Through the estimation of state-specific parameters, including location, scale, and skewness, the proposed model enables the detection of structural changes, such as shifts in the observed data distribution, while addressing challenges such as overfitting and computational inefficiencies inherent in Gaussian mixtures. Both simulation studies and real data analysis demonstrate the robustness and flexibility of the approach, highlighting its ability to accurately model asymmetric data and detect regime transitions. This methodological advancement broadens the applicability of a finite mixture of hidden Markov models across various fields, including demography, economics, finance, and environmental studies, offering a powerful tool for understanding complex temporal dynamics.  

\medskip



\setstretch{1.6}
\newpage

\section{Introduction}\label{Intro}

Finite Mixture Models (FMMs) have become indispensable in statistical modeling and data analysis due to their remarkable flexibility. These models serve two purposes: as a clustering technique and as a mathematical framework to derive analytically tractable forms of complex data distributions \citep[see, e.g.,][]{Titterington1985}. The extensive body of literature on FMMs highlights their applications in classification, clustering, density estimation, and pattern recognition, with foundational contributions from works such as \citet{Titterington1985, McLachlanBasord1988} and \citet{McLachlanPeel2000}, among others.

Mixture distributions have also proven to be highly effective in modeling heterogeneous populations, making them a valuable tool across a wide range of research fields. Applications span economics \citep{CompianiKitamura2016}, image processing \citep{Bohning2000}, cytometry \citep{LeeEtAl2016}, forestry \citep{JaworskiPodlaski2012}, social sciences \citep{ConnellFrye2006}, genetics \citep{JiaoZhang2008}, and medicine \citep{LinEtAl2007, LinEtAl2002}, to name only a few. For a comprehensive review of the applications of FMM distributions, see \citet{FruehwirthSchnatter2006} and \citet{ZhangHuang2015}.

The work of \citet{Pearson1894} is among the earliest to describe parameter estimation for FMMs using the method of moments. Although the most common approach for estimating FMM parameters is maximum likelihood estimation (MLE), it is not without challenges. Convergence issues can arise, as MLE relies on iterative techniques, such as the Newton-Raphson (NR) method, to numerically maximize the log-likelihood function. In such cases, NR methods may encounter local extrema, leading to a failure to find a root. Moreover, if the log-likelihood function is ill-posed, the NR method can enter an infinite loop, as observed by \citet{McLachlanPeel2000}. 

Advancements in computational methods, particularly Markov chain Monte Carlo (MCMC) techniques, have contributed significantly to overcoming these challenges. Bayesian approaches to mixture modeling have garnered considerable attention, as demonstrated by the influential works of \citet{DieboltRobert1994, EscobarWest1995, RichardsonGreen1997}, and \citet{Stephens2000}. In parallel, Expectation-Maximization (EM) algorithms have been developed as a powerful tool to address complex likelihood estimation problems, particularly in scenarios involving multiple optima. Although highly practical, these algorithms are not without limitations, including slow convergence, computational complexity, and challenges associated with (hyper-) parameter tuning.

Regardless of the optimization algorithm used, deviations from the normality assumption for mixture components can pose significant challenges, particularly when the data include observations with asymmetric distributions. In such cases, the selection of mixture components that can accommodate asymmetry offers a better fit. In many applied problems, the shapes of mixtures fitted with normal components may become distorted, leading to potentially misleading inferences when dealing with highly asymmetric data. Specifically, Gaussian mixture models tend to overfit by adding extra components to capture skewness. However, increasing the number of pseudo-components not only complicates interpretation but also introduces computational difficulties and inefficiencies.

To address these issues, we propose using skew-normal distributions, as introduced by \citet{Azzalini85}, to enhance the modeling of asymmetry. The seminal work of \citet{AzzaliniDallavalle} further extended this concept to the multivariate setting. In parallel, advances in handling skewness within distributions have led to the skew elliptical family and other related parametric families. Among these developments, the contributions of \citet{Azzalini2005, ValleAzzalini2006, Valleetal2006}, and \citet{Ma} have been particularly influential.

Since Azzalini's foundational contributions, this family of distributions has continued to attract significant attention in the scientific literature \citep[see, e.g.,][]{Arellano2022statpapers, wang2023Stat, WZW23}. The extension of the multivariate framework by \citet{AzzaliniDallavalle} has further stimulated research, generating additional outputs \citep[see, e.g.,][]{bufaloNigri2024}. Advancements in addressing skewness within distributions have paved the way for the development of the skew elliptical family and other parametric families. 

In recent decades, the development of parametric distribution families as alternatives to the normal distribution has gained significant momentum. This progress is driven by the availability of large multivariate data sets and advancements in computational power. These factors, combined with recent theoretical innovations, have enabled the creation of more realistic statistical models that extend beyond the traditional reliance on the normality assumption. 

Historically, the assumption of normality in statistical modeling was often adopted out of practical necessity rather than theoretical conviction, as it provided substantial simplifications in both theoretical derivations and computational processes. The inclusion of the Gaussian family within broader alternative frameworks remains conceptually appealing, aligning with the notion of a "departure from normality", which implies a gradual divergence from a standard or "normal" state. In particular, the Gaussian family often serves as a limiting case within these alternative families, preserving a weaker but meaningful connection to traditional models.

The introduction of skewed-normal distributions in longitudinal contexts enables a more precise representation of shape variations, particularly structural breaks, when the data exhibit significant asymmetries across different phases of a series. Change point detection plays a crucial role in time series analysis as it highlights significant shifts in the data-generating process. This technique finds applications in a wide range of fields, including economic and financial data analysis, environmental data monitoring, and image analysis, where segmentation is performed based on varying textures or objects. Notable contributions to this area include surveys by \citet{Burg2022} and \citet{Aminikhanghahi2017survey}, which provide a comprehensive overview of the subject.

The skewed-normal approach offers several significant advantages, including greater flexibility in modeling data distributions by capturing not only the mean and variance but also the asymmetry. This enables more accurate estimation of mixture components, even in the presence of structural changes with non-symmetric characteristics. Moreover, regime switches can be detected through variations in asymmetry patterns over time, providing enhanced robustness in identifying transitions between distributions and offering deeper insights into structural shifts. \citet{Ngunkeng2014} and \citet{Valle2013} are among the few who have effectively explored this topic using skew-normal distributions, demonstrating the benefits of an asymmetric approach. However, their proposed estimation methods differ significantly from our proposal documented in Section~\ref{sec:model}.

The appeal of the skewed distribution family lies in its remarkable flexibility, which allows it to accommodate a wide range of distributions and address numerous practical challenges in empirical data analysis. Our contribution seeks to further enhance this flexibility by introducing two significant methodological advancements:
\begin{inparaenum}[(i)]
\item We extend the skewed-normal distribution to mixture models, thereby enhancing its ability to capture complex data structures and asymmetries across multiple components.
\item We employ a two-state Hidden Markov Model (HMM), extended to a mixture of skewed-normal distributions, to effectively model location, scale, and shape parameters. This approach enables the identification and modeling of changes over time, offering a robust framework for capturing dynamic shifts in the data set.
\end{inparaenum}

As mentioned in point (ii), our proposal utilizes the HMM framework. An HMM is a practical approach to the analysis of time series data, where unobserved (hidden) states influence the behavior of observed time series \citep{baum1970maximization}. The fundamental work of \citet{rabiner1986introduction} and \citet{rabiner1989tutorial} provides a comprehensive understanding of HMMs and their broad applications. Recent advances have further expanded the scope of HMMs, including contaminated Gaussian HMMs \citep{punzo2016contaminated}, HMMs for matrix time series \citep{asilkalkan2021hidden}, and parsimonious HMMs for matrix-variate data \citep{tomarchio2021parsimonious}.

The combination of FMMs and HMMs has gained increasing popularity for modeling heterogeneous temporal data. Although FMMs help partition observations into distinct groups, HMMs capture homogeneous hidden states within these clusters. This powerful combination, known as a finite mixture of Hidden Markov Models (FMM-HMM), has found applications in various fields, including finance \citep{Dias2015, Knab2003model}, transportation \citep{chamroukhi2011hidden}, and clinical studies \citep{bartolucci2014latent, maruotti2012mixed}. Another approach involves using mixtures as emission distributions within HMM, which offers greater flexibility in modeling, including the ability to capture variable subsets over time \citep{volant2014flexible}.

Our study extends this framework by incorporating FMM-HMM. This model assumes a constant transition probability matrix over time, while the emission distribution varies according to the hidden state (regime), enabling the exploration of dynamic dependencies between variables. The paths derived from the estimated HMM can then be utilized to detect change points, which represent shifts in the observed distribution. This capability to identify structural changes within the framework adds an innovative element to our approach.

Unlike previous works that used skew-normal distributions in HMMs, our model explicitly combines finite mixture modeling with regime switching, allowing for state-dependent estimation of skewness, scale, and location parameters. This enables the capture of a broader range of structural changes and asymmetries in time series data. In fact, the existing literature proposes changepoint detection in skew-normal models, possibly using skew-normal emissions within HMMs. To the best of our knowledge, none of the existing approaches fully integrates the mixture model framework with a Markov switching process for dynamic changepoint analysis. This represents a further element of innovation in our work. In addition, the Bayesian implementation that we propose ensures robust detection of regime changes while promoting a tendency to avoid overfitting. This also represents an aspect that has been little explored in the literature.

\color{black}

The paper is structured as follows. Section~\ref{sec:model} introduces the methodological background and the proposed modeling approach. Section~\ref{sec:sim} presents a series of simulation studies, while a gender gap analysis in human mortality is presented in Section~\ref{sec:realdata}. Finally, Section~\ref{sec:conc} provides the discussion and conclusions, along with some ideas on how the methodology can be further extended.

\section{Model specification and methods}\label{sec:model}

Asymmetric density distributions are crucial in statistics for modeling phenomena where data exhibit non-symmetric patterns, such as when there is a pronounced accumulation toward one tail over the other. These distributions are particularly useful for modeling situations such as financial returns, meteorological data, environmental data, and other scenarios where asymmetry is a key feature.

Statistical references feature several asymmetric distributions, which are typically represented by density functions that incorporate skewness parameters, allowing for shapes that differ from those of symmetric distributions. One notable example is the skew-normal distribution introduced by \citet{Azzalini85, AzzaliniDallavalle} and \citet{Azzalini2005}, which extends the Gaussian distribution by adding a skewness parameter $\lambda$ to model longer tails on one side. The density function for the skew-normal distribution is given by the formula in~\eqref{SkN}, as follows:
\begin{equation}\label{SkN}
f(y; \xi, \omega^2, \lambda) = \frac{2}{\omega} \phi\left(\frac{y - \xi}{\omega}\right) \Phi\left(\lambda \frac{y - \xi}{\omega}\right),
\end{equation}
where the parameters $\xi$, $\omega$, and $\lambda$ are the location, scale, and shape parameters, respectively. The success of such distributions can be attributed to their tractability, the broad range of indices they can accommodate (including skewness and kurtosis), and the flexibility they offer in modeling a variety of shapes.

When considering asymmetry, the central role of the skewness parameter $\lambda$, becomes evident. When $\lambda = 0$, the skewness disappears and the model is simplified to a normal density. As $\lambda \to \infty$, the skewness intensifies, eventually converging to the half-normal (or folded normal) density function. Furthermore, if the sign of $\lambda$ changes, the density is reflected on the vertical axis, reversing the direction of skewness.

The model is further defined by $\phi(\cdot)$ and $\Phi(\cdot)$, which represent the standard normal probability density function and the cumulative distribution function, respectively. The first element, $\phi(\cdot)$, corresponds to:
\begin{equation*}
\phi(y) = \frac{1}{\sqrt{2 \pi}} e^{-\frac{y^2}{2}}.
\end{equation*}
and describes the probability of observing a certain value \(y\) in a standard normal distribution (mean \(0\) and variance \(1\)), while the standard normal cumulative distribution function is formally represented as
\begin{equation*}
\Phi(y) = \int_{-\infty}^{y} \phi(t) \, dt,
\end{equation*}
where \(\Phi(\cdot)\) represents the probability that a standard normal random variable takes a value less than or equal to \(y\), thus returning the cumulative value of the standard normal distribution up to \(y\). In the skew-normal distribution, \(\Phi\left(\lambda \frac{y - \xi}{\omega}\right)\) introduces asymmetry by adjusting the cumulative probability based on the skewness parameter \(\lambda\), thus modifying the shape of the distribution.

\subsection{Finite mixture models}\label{Sec:FMM}

Incorporating asymmetric processes helps address the need to model data that can not be adequately represented by overly simplistic distributions. However, this approach does not fully capture the complexities encountered in empirical data, particularly when the observation horizon extends and time series may experience one or more structural breaks. FMMs were therefore developed to provide computationally useful representations for modeling complex data distributions over random phenomena \citep[see, e.g.,][]{McLachlan2019}.

In general terms, FMMs represent a broad class of statistical models that can be used for inference problems involving heterogeneous values and for clustering units based on unobserved characteristics \citep{Aitkin1980mixture}. The underlying concept of these models is that the data may originate from two or more latent clusters, each potentially sharing common distribution shapes but differing in their parameters \citep{AR1985}.

A mixture model is a collection of distributions or probability densities represented as a weighted sum (with weights \(\pi\)) of the component densities (\(k\)) that make up the mixture. The term "finite" indicates that the component distributions are defined by a finite number of parameters. Although the model can be extended to include an infinite number of components, in practice, it is sufficient to work with finite mixtures, as an infinite mixture of distributions can be effectively approximated by a finite number of components.

In general terms, a mixture can be formalized as follows:
\begin{equation}\label{General_Mixture}
f_{(y;\Theta_1 \dots \Theta_k)} = \sum_{k=1}^K \zeta_k g_k (y;\Theta_k)
\end{equation}
with the condition that $\sum_{k=1}^{K} \zeta_k = 1$.

The most commonly used finite mixture distributions involve Gaussian components \citep{Everitt2013}; however, it is easy to envision adaptations for our case, such as considering a mixture of two components, each with a skew-normal ($\mathcal{SN}$) density. In this setup, we have \(\pi_1 + \pi_2 = 1\), which ensures that the basic properties of a probability density function are satisfied. This is formalized as follows:
\begin{equation*}
\zeta_1 \times  \mathcal{SN}_1\left(y_1 ; \xi_1, \omega_1^2, \lambda_1\right) 
+\zeta_2  \times  \mathcal{SN}_2\left(y_2 ; \xi_2, \omega_2^2, \lambda_2\right).
\end{equation*}

Although our approach is general and applicable in various contexts, the focus of this article is on defining multiple components within a time series. In this regard, the introduction of mixtures facilitates the detection of potential structural breaks, while extending the model to include asymmetry allows for the estimation of regime changes based not only on mean and variance, but also on higher-order distribution moments\footnote{It is worth providing a simple way to obtain higher moments without using the moment-generating function. With some basic algebraic manipulations, we can easily obtain
$E(Y)=\xi+\sqrt{\frac{2}{\pi}} \delta(\lambda) \omega $,
$\operatorname{var}(Y)=\left\{1-\frac{2}{\pi} \delta^2(\lambda)\right\} \omega^2, $,
$\gamma_Y=\frac{\sqrt{2}(4-\pi) \lambda^3}{\left\{\pi+(\pi-2) \lambda^2\right\}^{3 / 2}}$, $\kappa_Y=3+\frac{8(\pi-3) \lambda^4}{\left\{\pi+(\pi-2) \lambda^2\right\}^2}$, where \(\delta(\lambda)=\lambda / \sqrt{1+\lambda^2}\), and \(\gamma_Y\) and $\kappa_Y$ are measures of skewness and kurtosis, respectively.}.

The use of asymmetric distributions enhances the model's ability to distinguish transitions between regimes, reducing the risk of ``false positives" caused by symmetric oscillations around the mean. This facilitates the accurate identification of true breakpoints and strengthens the robustness of the approach. In addition, it improves the interpretation of structural changes, providing a more nuanced understanding of the underlying phenomenon.

To meet the specifics of a longitudinal data approach, HMMs and Viterbi algorithms have been implemented, the details of which will be presented in Sections~\ref{sec:HMM} and~\ref{sec:2.3}.

\subsection{Hidden Markov models}\label{sec:HMM}

The theory of HMMs is introduced. These include:
\begin{inparaenum} 
\item[1)] The general formulation and properties of HMMs are described. 
\item[2)] Methods for estimating hidden Markov chains are discussed, including an introduction to filtering and smoothing techniques. 
\item[3)] The Viterbi algorithm is presented as a computational method to determine the most probable state sequence. 
\end{inparaenum}

A HMM is a model in which a sequence of emissions (or outputs) is observed, but this sequence is modeled through a latent (i.e., hidden) state sequence, which is assumed to follow a Markov chain. The objective of analyzing HMMs is to infer the underlying sequence of hidden states from the observed data.

The mathematical formulation of HMMs was first proposed by \cite{Baum1966}, while only in \cite{rabiner1989tutorial} a first practical use in speech recognition can be traced. Nowadays, HMMs may be found in various fields of application and are commonly used for different types of time series modeling, where the observed sequence can be complemented by a latent Markov chain.

HMMs derive their name from two key properties. The first property is the assumption that an observation $y_t$, for $t = 1, \ldots, T$, is generated by an underlying process whose state is hidden and unobservable. Let there be \( Z \) distinct states, with $z_t = 1, 2, \ldots, Z$ representing the state at time $t$. The second property is that the state follows a discrete-time (first-order) Markov process, which implies that the probability of transitioning to the current state depends only on the previous state. Expressing it mathematically
\begin{equation*}
p\left(z_t \mid z_{t-1}, z_{t-2}, \ldots, z_1\right) \equiv p\left(z_t \mid z_{t-1}\right).
\end{equation*}

A stochastic process is considered a Markov process if, given the value of \( z_{t-1} \), the current state \( z_t \) is independent of all previous states (those prior to \( t-1 \)). The HMM also satisfies the Markov property with respect to the observations \( y_t \). Specifically, given \( z_t \), the observation \( y_t \) is independent of all states and observations at previous times, ensuring that the sequence of observations is conditionally independent once the hidden state is known.

Let \( \mathbf{y}_{1:T} = \{y_t\}_{t=1}^T \) represent the sequence of observed variables indexed by time \( t = 1, \ldots, T \), and let \( \mathbf{z}_{1:T} = \{z_t\}_{t=1}^T \) represent the sequence of hidden states. The bivariate process \( \{y_t, z_t\}_{t=1}^T \) is thus referred to as a HMM. Following the general form introduced by \cite{DK12}, this process is defined as:
\begin{equation*}
y_t \sim p\left(y_t \mid z_t\right), \quad z_t \sim p\left(z_t \mid z_{t-1}\right), \quad z_1 \sim s_i=p\left(z_1=i\right), \quad 1 \leq i \leq Z .
\end{equation*}

The transition matrix $\mathbf{A}=\left\{a_{i j}\right\}_{1 \leq i, j \leq Z}$ is a $Z \times Z$ matrix of state transition probabilities, where $a_{i j}=p\left(z_t=j \mid z_{t-1}=i\right)$. The state transition coefficients have the properties $\sum_{i=1}^Z a_{i j}=1$ and $a_{i j} \geq 0; \boldsymbol{s}=\left\{s_i\right\}_{i=1}^Z$ is a vector with the initial state probability distributions.

In the HMM literature, two functions are commonly referred to as likelihood. The first is the so-called complete-data likelihood, which represents the joint distribution of the observations \( \{y_t, z_t\}_{t=1}^T \). This is the probability that both \( \mathbf{y}_{1:T} \) and \( \mathbf{z}_{1:T} \) co-occur. The complete specification of the likelihood can be formalized as follows: 
\begin{align*}
\mathcal{L} & =p\left(\mathbf{y}_{1: T}, \mathbf{z}_{1: T}\right)=p\left(\mathbf{z}_{1: T}\right) p\left(\mathbf{y}_{1: T} \mid \mathbf{z}_{1: T}\right) \\
& =\underbrace{s_i \prod_{t=2}^T p\left(z_t \mid z_{t-1}\right)}_{\text {latent Markov process }} \underbrace{\prod_{t=1}^T p\left(y_t \mid z_t\right)}_{\text {observations }} \\
& =s_i \prod_{t=2}^T a_{z_{t-1}, z_t} \prod_{t=1}^T p\left(y_t \mid z_t\right) .
\end{align*}

The second type of likelihood is the marginal likelihood, which requires summation of all possible state sequences \citep{LM18}. This can be expressed as follows:
\begin{equation*}
\mathcal{L}_m=p\left(\mathbf{y}_{1: T}\right)=\sum_{z_1=1}^K \ldots \sum_{z_T=1}^K s_i \prod_{t=2}^T a_{z_{t-1}, z_t} \prod_{t=1}^T p\left(y_t \mid z_t\right) .
\end{equation*}

The state-dependent emission distribution, $p\left(y_t \mid z_t\right)$, can be discrete or continuous in HMMs, where each observation $y_t$ is generated from state $z_t$. Since we have set up a mixture composed of two skew-normal distribution components, the joint distribution of the hidden Markov process is given by:
\begin{equation*}
p\left(\mathbf{z}_{1: T}, \mathbf{y}_{1: T}\right)=p\left(\mathbf{z}_{1: T}\right) p\left(\mathbf{y}_{1: T} \mid \mathbf{z}_{1: T}\right)=\left[p\left(z_1\right) \prod_{t=2}^T p\left(z_t \mid z_{t-1}\right)\right]\left[\prod_{t=1}^T p\left(\mathbf{y}_t \mid z_t\right)\right],
\end{equation*}
where \( p(\mathbf{z}_1) \) represents the distribution of the initial state, specifying the probabilities of the initial hidden state. The transitions \( p(z_t | z_{t-1}) \) capture the dynamics of the hidden states over time, while the probabilities \( p(\mathbf{y}_t | z_t) \) define the likelihood of observing the data \( \mathbf{y}_t \) given the hidden state \( z_t \), with the observations modeled as a mixture of skew-normal distributions.

Thus, formula~\eqref{General_Mixture} can hence be rewritten in terms of probabilities conditioned on the observed states as follows:
\begin{equation*}
p\left(\mathbf{y}_t \mid z_t, \boldsymbol{\Theta}\right)=\sum_{k=1}^K \zeta_k g_k\left(\mathbf{y}_t \mid \boldsymbol{\Theta}\right),
\end{equation*}
where $g_k$ is the $k$\textsuperscript{th} density function of a skew-normal distribution $\mathcal{SN}\left(\xi_k, \omega_k^2, \lambda_k\right)$ with $\xi$, $\omega$, and $\lambda$ are, as above, the location, scale, and shape parameters of the $k$\textsuperscript{th} density and
 \(\boldsymbol{\Theta} \in \mathbb{R}^{k\times Z}\), while the general transition for matrix $\mathbf{A}$ with dimensions $Z \times Z$ defined by:
\begin{equation*}
\mathbf{A}=\left(\begin{array}{cccc}
a_{11} & a_{12} & \cdots & a_{1 Z} \\
a_{21} & a_{22} & \cdots & a_{2 Z} \\
\vdots & \vdots & \ddots & \vdots \\
a_{Z 1} & a_{Z 2} & \cdots & a_{Z Z}
\end{array}\right),
\end{equation*}
which is subject to the row-sum constraint (that is, $\sum_{j=1}^Z a_{i j}=1$, for each $i \in\{1, \ldots, Z\}$). Following this notation, a first-order Markov process will therefore be subject to the following conditional probability: $a_{i j}=p\left(z_t=j \mid z_{t-1}=i\right)$.

\begin{figure}[!htb]
\centering
\includegraphics[width=1\linewidth]{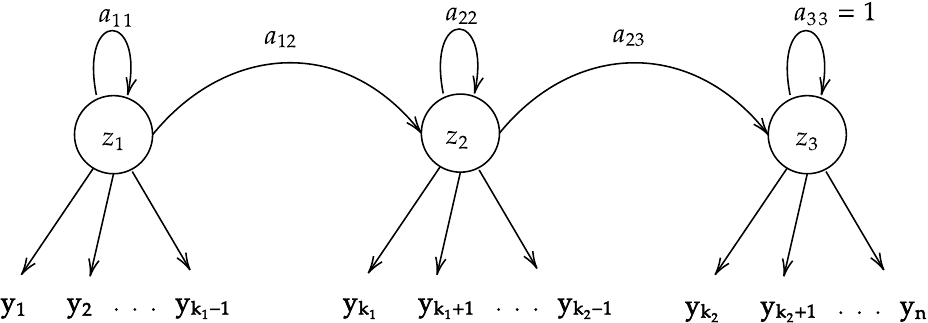}
\caption{\small View of a HMM model with three states and a change points.}\label{fig:hmm}
\end{figure}

Based on the Markovian property described, Figure~\ref{fig:hmm} presents a schematic representation of the three-state HMM model. By combining the Markov-switching structure with a mixture of skew-normal distributions, this approach effectively addresses the question of ``if and where a change in the outcome has occurred." In this context, the most probable sequence and any structural changes in the observed data can be estimated using the Viterbi algorithm, which is briefly outlined in the following subsection.

\paragraph{Prior Distributions}

Leveraging the Bayesian framework for the proposed model, the prior distributions are specified as follows:

For each row of the transition probability matrix \(\mathbf{A}\), and the mixing proportions \(\zeta\), we adopt a Dirichlet prior: \(A_{i.}, \zeta \sim \text{Dirichlet}(\mathbf{1})\), where \(\mathbf{1}\) denotes a vector of ones, reflecting the same prior probability across states and mixtures.

The means of the skew-normal distributions, \(\xi_1\) and \(\xi_2\), are assigned a normal prior: \(\xi_1, \xi_2 \sim \mathcal{N}(0, 10)\). The scale parameters, \(\omega_1\) and \(\omega_2\), are given weakly informative priors to promote flexibility while avoiding extreme values: \(\omega_1, \omega_2 \sim \text{Cauchy}(0, 2)\). The skewness parameters, \(\lambda_1\) and \(\lambda_2\), are assumed to follow a standard Gaussian distribution: \(\lambda_1, \lambda_2 \sim \mathcal{N}(0, 1)\).
Lastly, the expectation of $a_{i j}$, a priori, is 0.9 with a standard deviation of approximately 0.07. This should guarantee obtaining a posterior that properly assesses whether a jump occurred. This is possible because of our framework that will not easily jump from state $i$ to state $i+1$ unless there is strong evidence. Although this prior can be considered subjective, as often occurs in Bayesian modeling, our choice is oriented to a fairly non-informative approach in the sense that our coefficients are on an exponential scale. 

\subsection{State and parameter estimation methods}\label{sec:2.3}

Our proposal aims to estimate the parameters of a skew-normal mixture modeled with two states, enabling the identification of potential changepoints. In this framework, the hidden Markov processes described above are integrated within a Bayesian perspective to effectively address the problem, enhancing the accuracy of the parameter estimates. Given the data sequence \(\mathbf{y}_{1: T}\), the approach can be formalized as follows:
\begin{equation*}
p\left(\boldsymbol{\theta} \mid \mathbf{y}_{1: T}\right)=\frac{p\left(\mathbf{y}_{1: T} \mid \boldsymbol{\theta}\right) p(\boldsymbol{\theta})}{\int_{\boldsymbol{\theta}} p\left(\mathbf{y}_{1: T} \mid \boldsymbol{\theta}\right) p(\boldsymbol{\theta}) d \boldsymbol{\theta}},
\end{equation*}
where the posterior probability, indicated on the left-hand side of the equation, is equal to the product of the marginal likelihood \(p\left(\mathbf{y}_{1: T} \mid \boldsymbol{\theta}\right)\) and the prior distribution of the parameters \(p(\boldsymbol{\theta})\), with the denominator acting as a normalization constant, which ensures that the posterior density integrates into one.

In the context of Bayesian estimation for HMM processes to identify changepoints in FMMs, techniques like Gibbs sampling and Metropolis-Hastings are often integral to various stages of the algorithm, particularly for the iterative estimation of parameters and inference on hidden states. Gibbs sampling is particularly useful when conditional distributions are known or easy to sample. On the other hand, Metropolis-Hastings is preferred in situations where the conditional distributions are not available in closed form or cannot be directly sampled. In some variants of HMMs, where hidden states exhibit complex dynamics or non-trivial dependencies, random-walk Metropolis-Hastings can be employed to sample entire segments of hidden states.

In situations where direct parameter estimation is infeasible or computationally inefficient, MCMC sampling can be used to simulate the HMM parameters from their posterior distribution, given a set of observed sequences. In this approach, the MCMC sampler alternates between the phases of sampling the parameters conditioned on the data and the hidden Markov chain and the phases of updating the HMM conditioned on the data and the parameters. This iterative process allows for the efficient estimation of the parameters and hidden states, even in complex models.

A particular class of MCMC methods is the Hamiltonian Monte Carlo (HMC), introduced to overcome the slow convergence of traditional techniques, especially in the context of complex models. The slow convergence in these models is often caused by the ``random walk" behavior of the processes being studied. In contrast, HMC mitigates this issue by introducing auxiliary momentum variables ($r$) for each model parameter ($\boldsymbol{\theta}$), allowing for a more efficient and faster exploration of the target distribution. This approach enhances the overall sampling efficiency, making it particularly useful for complex models with high-dimensional parameter spaces.

In the procedures included in RStan for computing HMC, the momentum variables $r$ are sampled independently of a reference distribution, where the joint density of $\boldsymbol{\theta}$ and $r$, $p(\boldsymbol{\theta}, r)$, defines the Hamiltonian dynamic system $H(\boldsymbol{\theta}, r)$, which is expressed as:
\begin{equation*}
\begin{aligned}
H(\boldsymbol{\theta}, r) & =-\log p(\boldsymbol{\theta}, r) \\
& =-\log p(r \mid \boldsymbol{\theta})-\log p(\boldsymbol{\theta}) \\
& =K(r \mid \boldsymbol{\theta})+V(\boldsymbol{\theta}),
\end{aligned}
\end{equation*}
where $K(r \mid \boldsymbol{\theta})$ represents the term associated with the density of the auxiliary momentum and $V(\boldsymbol{\theta})$ represents the term associated with the density of the target distribution. For a more comprehensive discussion of the technical details, refer to \cite{neal2012hmc}.

Once the system is defined, the HMC algorithm operates iteratively in two steps, where $r$ is updated first, followed by an update of~$\boldsymbol{\theta}$, using Hamilton’s equations for time \( t \), formulated as follows:
\begin{equation*}
\begin{aligned}
& \frac{d \boldsymbol{\theta}}{d t}=+\frac{\partial H}{\partial r}=\frac{\partial K}{\partial r}, \\
& \frac{d r}{d t}=-\frac{\partial H}{\partial \boldsymbol{\theta}}=-\frac{\partial K}{\partial \boldsymbol{\theta}}-\frac{\partial V}{\partial \boldsymbol{\theta}},
\end{aligned}
\end{equation*}
where \( \frac{\partial V}{\partial \boldsymbol{\theta}} \) is the gradient of the logarithm of the target distribution. The Hamiltonian dynamics is simulated for \( L \) steps, often using the \textit{"leapfrog"} method, which works as follows:
\begin{equation*}
\begin{aligned}
r(n+\epsilon / 2) & =r(n)-(\epsilon/2) \frac{\partial V}{\partial \boldsymbol{\theta}}[\boldsymbol{\theta}(n)], \\
\boldsymbol{\theta}(n+\epsilon) & =\boldsymbol{\theta}(n)+\epsilon \frac{r(n+\epsilon / 2)}{m}, \\
r(n+\epsilon) & =r(n+\epsilon / 2)-(\epsilon/2) \frac{\partial V}{\partial \boldsymbol{\theta}}[\boldsymbol{\theta}(n+\epsilon)],
\end{aligned}
\end{equation*}
where $n$ is the time, $\epsilon$ is the step size and $m$ is the mass associated with $\boldsymbol{\theta}$.

Then a new state $(\boldsymbol{\theta}^*, r^{*})$ is proposed and adopted only if accepted with probability:
\begin{equation*}
\alpha =\min \left\{1, \exp^{\left[-H\left(\boldsymbol{\theta}^*, r^*\right)+H(\boldsymbol{\theta}, r)\right]}\right\}.
\end{equation*}

The strength of this approach lies in the fact that the jumps made during the exploration of the parameter space are not random, but are instead guided by specific trajectories that are linked to the density. This feature enhances the efficiency of the algorithm, reducing the number of iterations required for convergence.

In the analytical framework described, HMC is complemented by a Viterbi-type algorithm. Although the estimation techniques outlined are well-established and widely recognized in the literature, they may struggle to correctly assign observations to their corresponding states, ultimately affecting the accuracy of parameter estimates. To address this challenge, \citeauthor{viterbi1967}'s \citeyearpar{viterbi1967} algorithm offers an optimal recursive solution for estimating the finite sequence of states in a discrete-time Markov process. By maximizing the posterior, it provides the most probable sequence of states, known as the Viterbi path, which is especially useful in situations involving a sequence of observed events.

While Markovian properties make parameter estimation problems tractable, the Viterbi algorithm enhances performance by enabling an efficient exhaustive search and excluding certain combinations from the problem. In practical terms, given a set of observations, the algorithm alternates between forward and backward phases to determine the most probable sequence of states (MAP path) in an HMM. The forward phase recursively computes the highest probability of reaching each state \(j\) at time \(t\), given that the most probable path has been followed up to that point. Thus, initializing at time \(t = 1\), for each state \(j\), the initial probability is calculated as:
\begin{equation*}
\delta_1(j) = s_j p(y_1 \mid z_1 = j),
\end{equation*}
where \(s_j\) is the initial probability of being in state \(j\), and \(p(y_1 \mid z_1 = j)\) is the probability of observing \(y_1\) in state \(j\). At each subsequent time \(t > 1\), for each state \(j\) at time \(t\), the probability of arriving there is calculated by starting from any state \(i\) at the previous time (\(t-1\)) and following the most probable path:
\begin{equation*}
\delta_t(j) = \max_i \left[ \delta_{t-1}(i) a_{ij} \right] p(y_t \mid z_t = j),
\end{equation*}
where \(\delta_{t-1}(i)\) is the highest probability of reaching state \(i\) at time \(t-1\), \(a_{ij}\) is the probability of transition from state \(i\) to state \(j\), and \(p(y_t \mid z_t = j)\) is the probability of observation \(y_t\) in state \(j\).

During the forward phase, for each state \(j\) at time \(t\), the previous state \(i\) that maximized the probability is also stored, creating a pointer for the backward trace phase. The reconstruction of the most probable sequence of states starts from the final state and works backward. The final state \(z_T^*\) at time \(T\) with the maximum probability is calculated as:
\begin{equation*}
z_T^* = \underset{i}{\arg \max} \ \delta_T(i),
\end{equation*}
where \(\delta_T(i)\) is the highest probability of being in state \(i\) at time \(T\). For each \(t = T-1, \dots, 1\), the most probable sequence of states is determined following the stored pointers, tracing back to the preceding state that maximized the probability at each step:
\begin{equation*}
z_t^* = \text{arg max}_{i} \left[ \delta_{t-1}(i) a_{i z_{t+1}^*} \right].
\end{equation*}

In other words, the state at time \(t\) is determined on the basis of the next state \(z_{t+1}^*\) and the saved pointer. This process makes the Viterbi algorithm effective in determining the most probable sequence of states for a given set of observations.

\section{Simulation studies}\label{sec:sim}

We present and discuss the results of a simulation study designed to evaluate the performance of the proposed finite-mixture HMM with skew-normal components. The analysis is conducted in two scenarios: a two-state model and a three-state model. These configurations enable us to assess the model's effectiveness in recovering parameters, detecting regime changes, and accurately classifying observations into their respective hidden states.

\subsection{Two-state model}

\begin{figure}[!htb]
\centering
\includegraphics[width=0.63\linewidth]{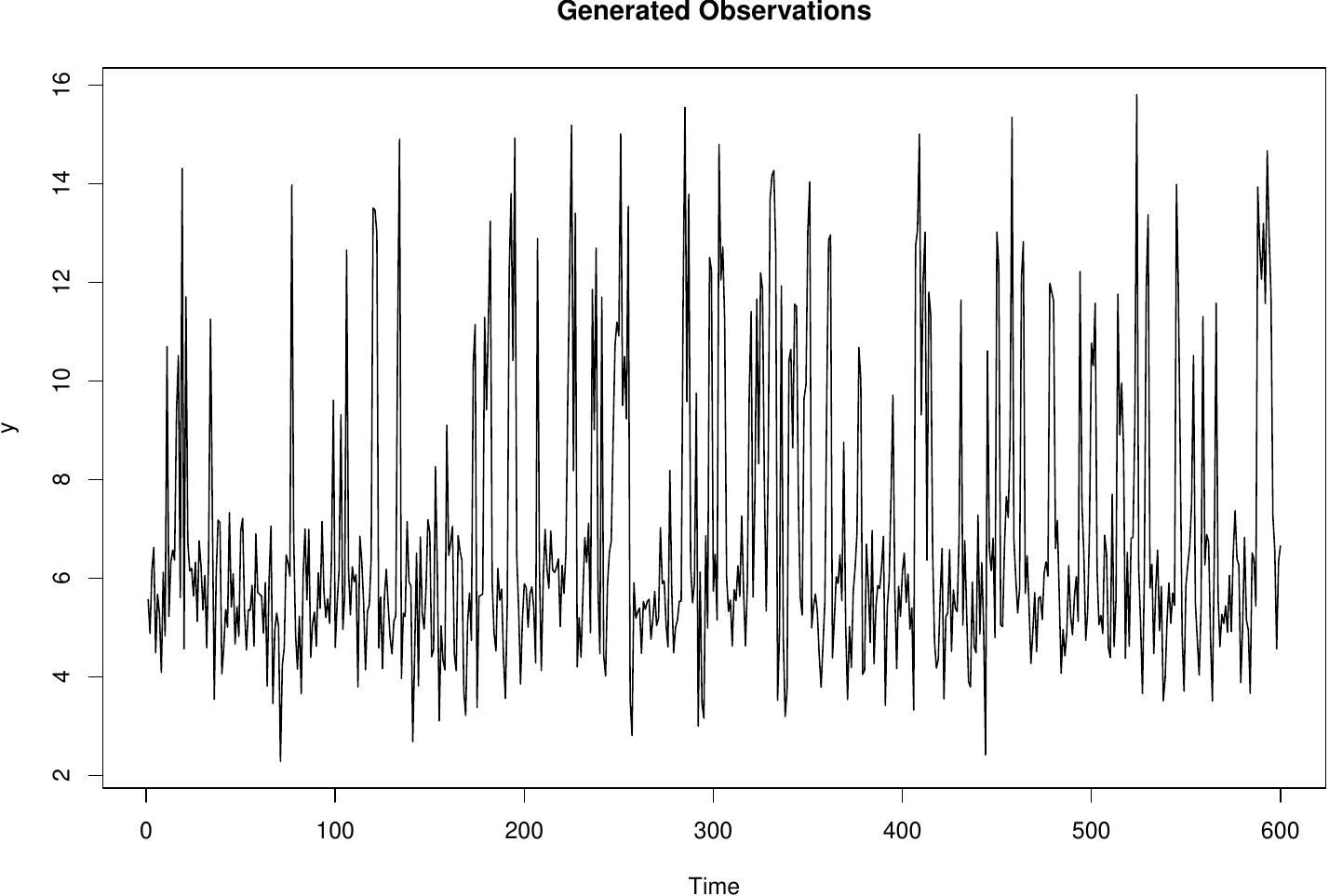}
\caption{\small Generated observations (two states)}\label{fig:Gen2}
\end{figure}

A comprehensive simulation study has been conducted to assess the performance of the proposed method. The study involves generating 600 data samples with two states, as illustrated in Figure~\ref{fig:Gen2}.

Recall that $\xi$, $\omega$, and $\lambda$ represents the location, scale, and shape parameters of the $k$\textsuperscript{th} density, the simulation setup is summarized in Table~\ref{tab:sim1}, with the transition probabilities simulated using the matrix defined in~\eqref{eq:A2}.


\begin{table}[!htb]
\centering
\caption{\small Generated Parameters for two states}
\centering
\begin{tabular}[t]{@{}lrrr@{}}
\toprule
Parameter & State 1 & Parameter & State 2 \\
\midrule
$\xi_1$ & 5.6328 & $\xi_1$ & 10.4042 \\
$\xi_2$ & 4.8938 & $\xi_2$ & 11.5115 \\
\midrule
$\omega^2_1$ & 0.9526 & $\omega^2_1$ & 2.0092 \\
$\omega^2_2$ & 0.9686 & $\omega^2_2$ & 1.6524 \\
\midrule
$\lambda_1$ & 0.9777 & $\lambda_1$ & 0.8933 \\
$\lambda_2$ & -0.8351 & $\lambda_2$ & 0.0284 \\
\midrule
$\zeta$ & 0.9048 & $\zeta$ & 0.0951 \\
\bottomrule
\end{tabular}
\label{tab:sim1}
\end{table}

\paragraph{Results of the two-state model}

\begin{figure}[!htb]
\centering
\includegraphics[width=0.46\linewidth]{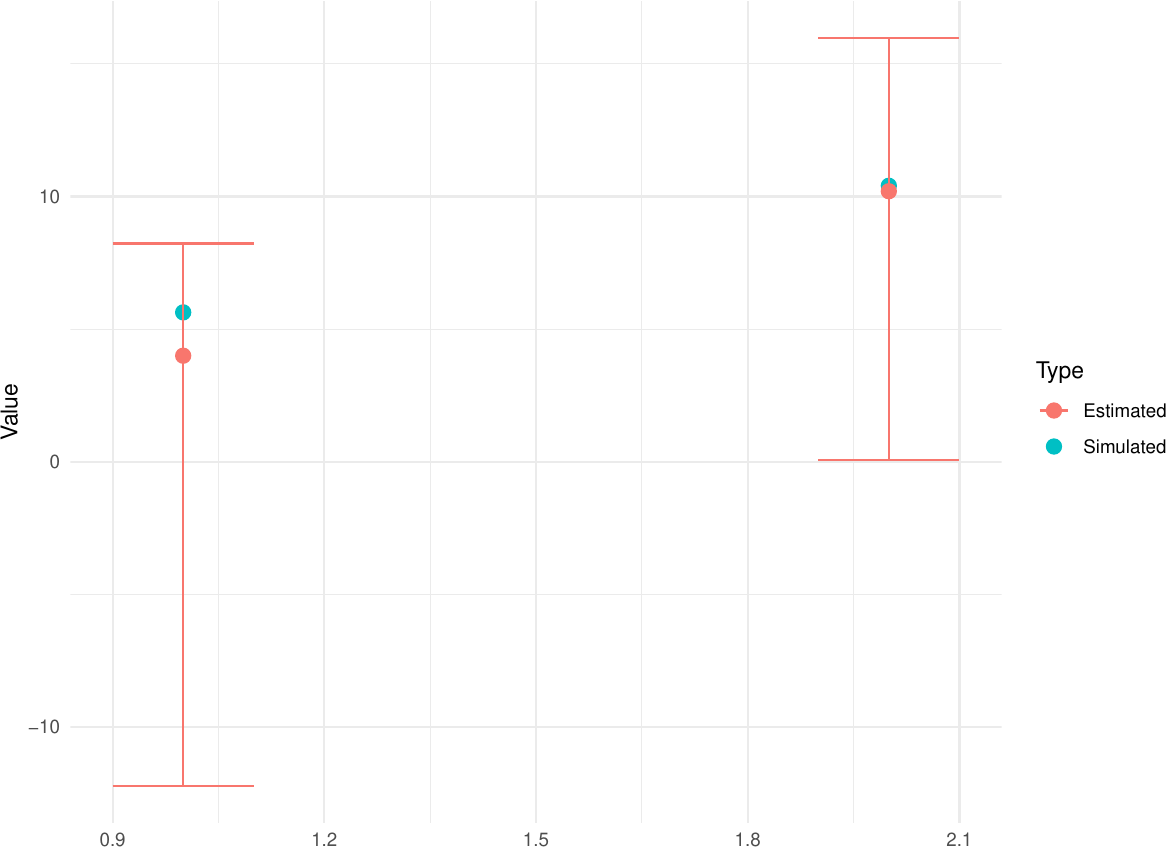}
\includegraphics[width=0.46\linewidth]{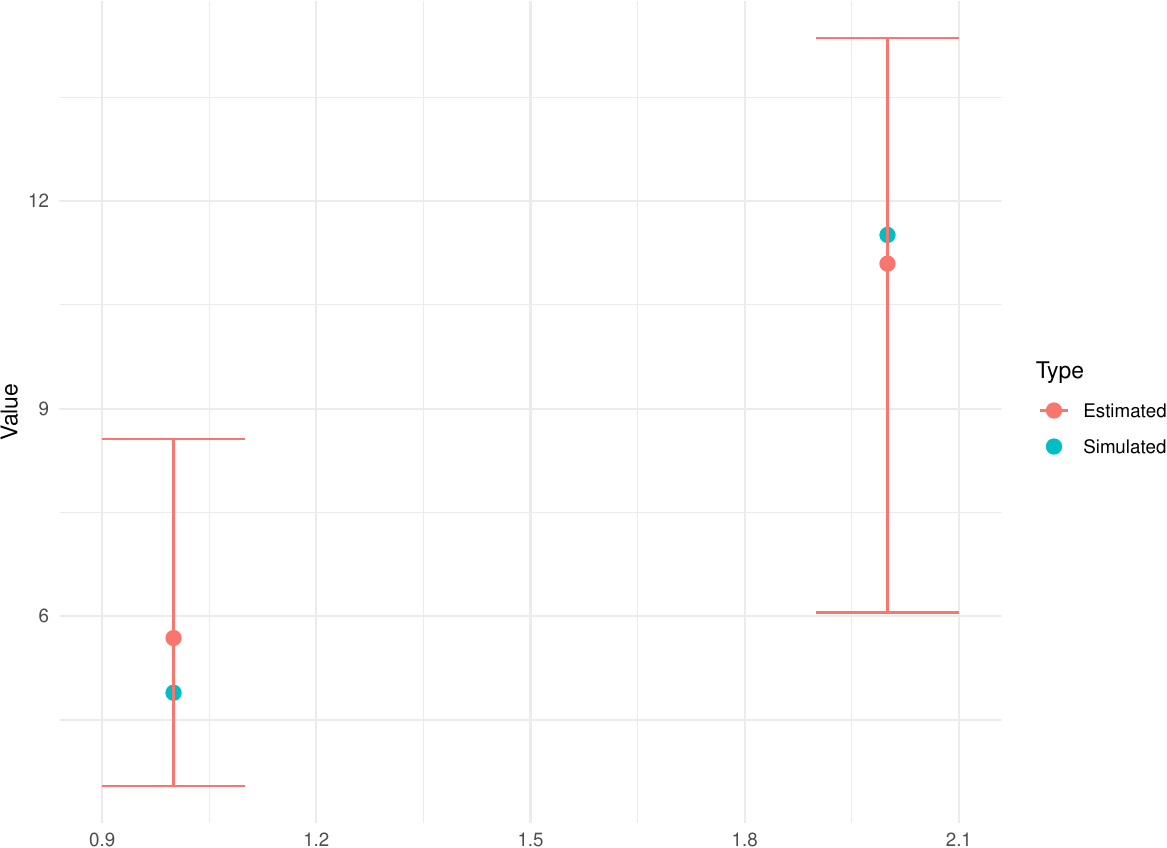}
\par\vspace{0.5cm} 
\includegraphics[width=0.46\linewidth]{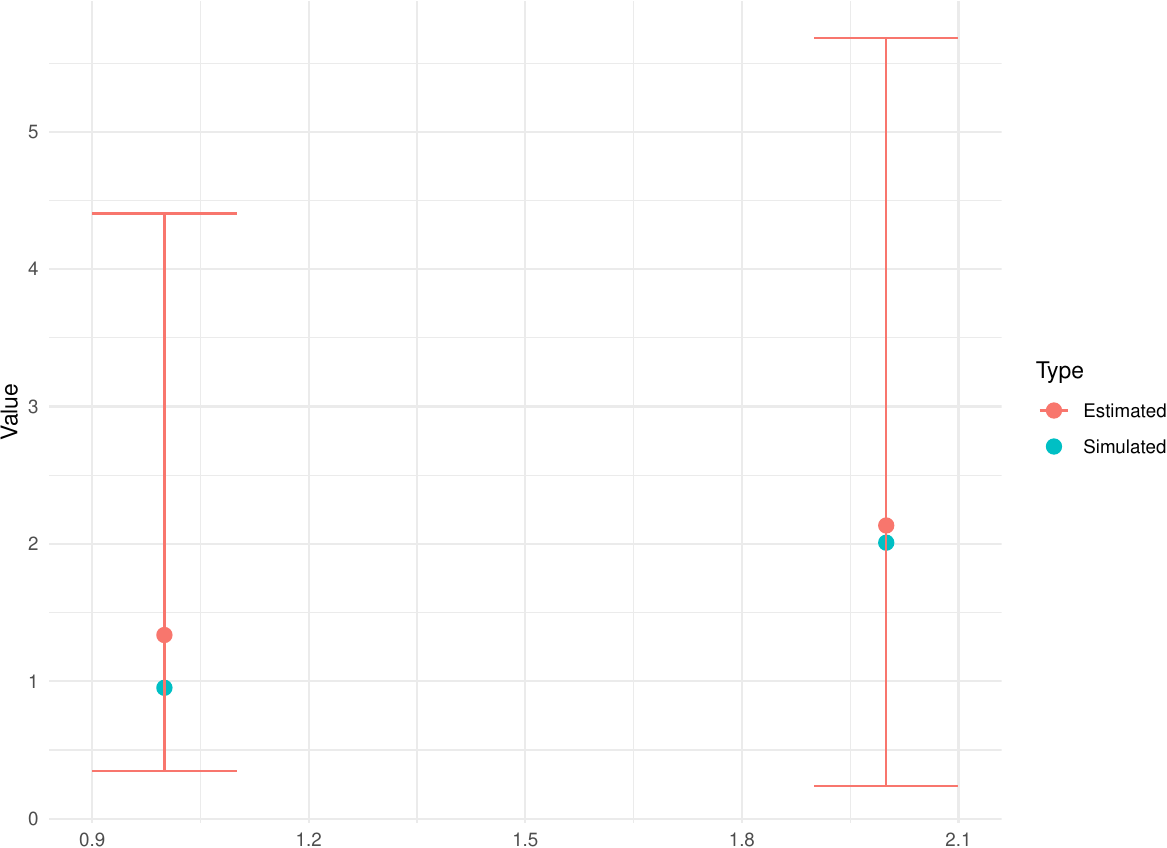}
\includegraphics[width=0.46\linewidth]{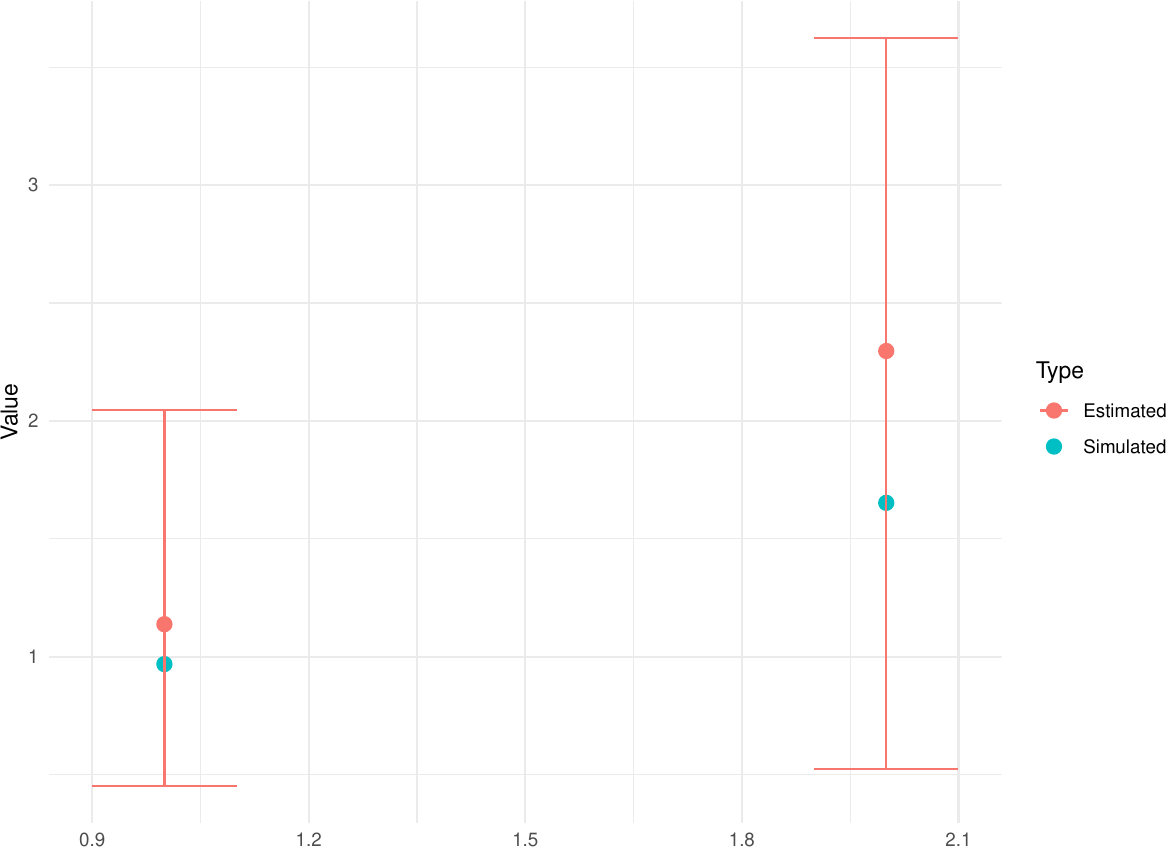}
\par\vspace{0.5cm} 
\includegraphics[width=0.46\linewidth]{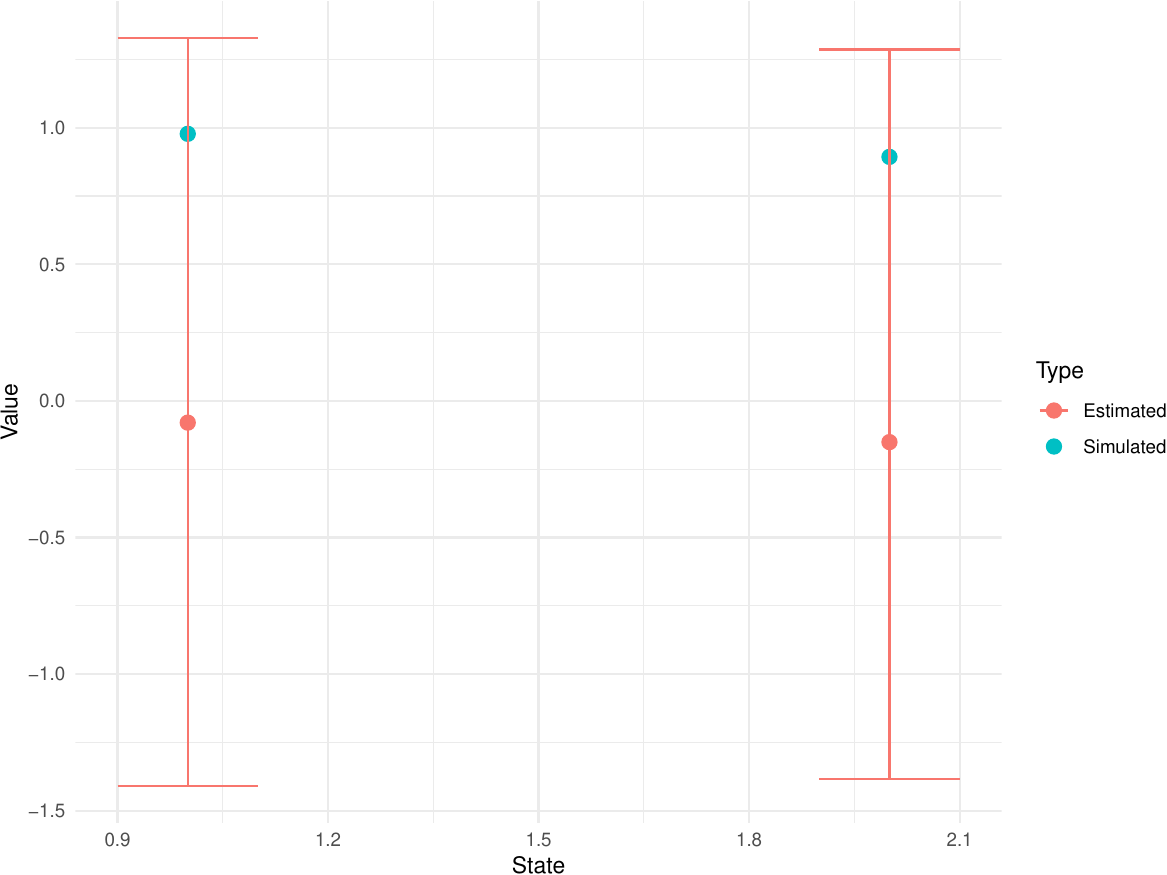}
\includegraphics[width=0.46\linewidth]{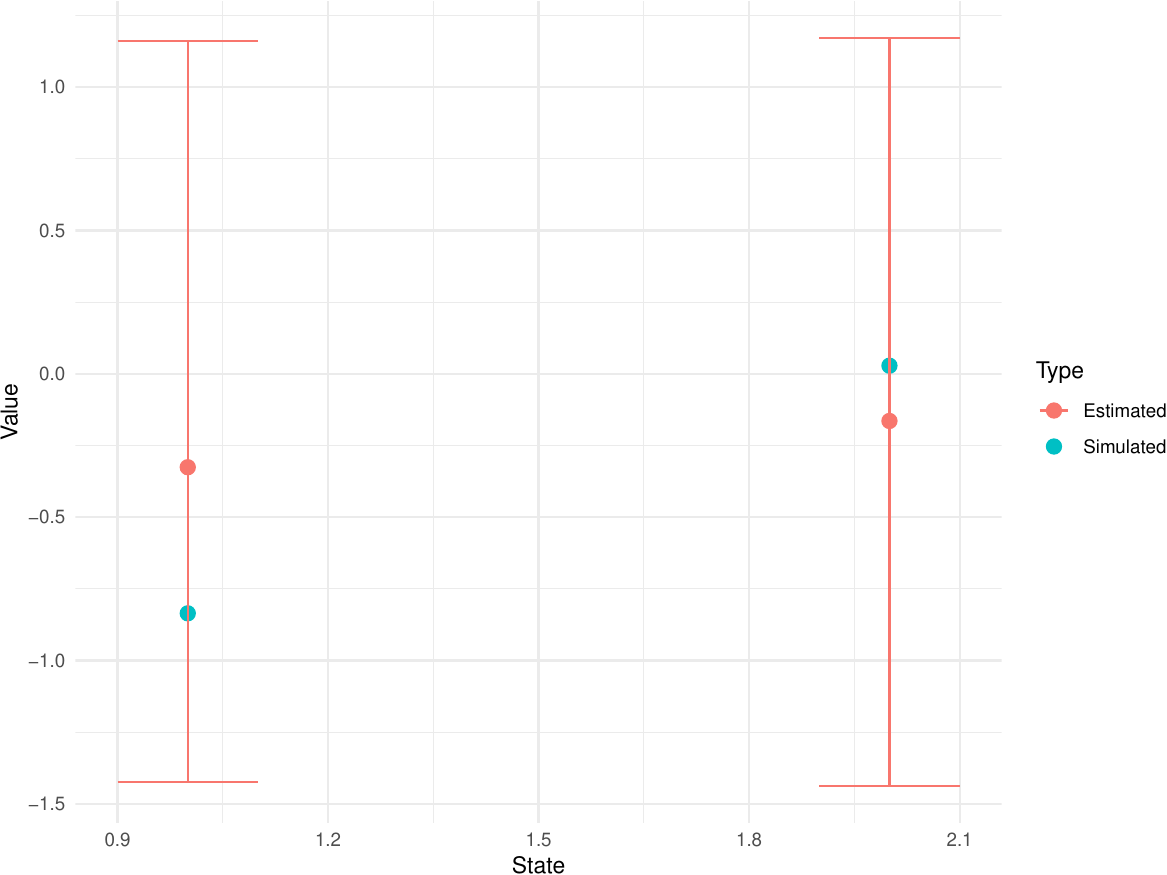}
\caption{\small Comparison between simulated and estimated values for location, scale, and skewness. Top row: location values. Middle row: scale values. Bottom row: skewness values.}
\label{fig:comparison2}
\end{figure}

The two-state simulation represents a relatively simple yet informative scenario for testing the robustness of the proposed methodology. The results indicate that the model successfully recovers the parameters used to generate the data. Specifically, we can observe in Figure~\ref{fig:comparison2} how the estimated parameters for location, scale, and skewness closely match the simulated values for both hidden states. 


    

The estimated skewness parameters, which play a critical role in capturing the asymmetric nature of the observations, may differ slightly from the expectations for the first state. However, this discrepancy does not undermine the analysis as the overall alignment remains consistent, confirming the model's ability to capture data asymmetry and regime-specific characteristics. Ultimately, the model delivers accurate estimations, highlighting the effectiveness of incorporating skew-normal components into the HMM framework.

This alignment reinforces the model's capacity to account for data asymmetry and regime-specific characteristics. Specifically, the skewness parameter is essential for capturing the asymmetric nature of the observations, and its accurate estimation underscores the effectiveness of incorporating skew-normal components into the HMM framework.
\begin{equation}
\mathbf{A}=\left(\begin{array}{cccc}
0.8707 & 0.1292\\
0.4035 & 0.5964\\
\end{array}\right).
\label{eq:A2}
\end{equation}

The aforementioned estimated probability transition matrix \(\textbf{A}\) closely reflects the underlying transition dynamics between the two states. The high diagonal probabilities indicate the persistence of each state, whereas the off-diagonal elements capture the infrequent shifts between regimes.

\begin{figure}[!htb]
\centering
\includegraphics[width=0.61\linewidth]{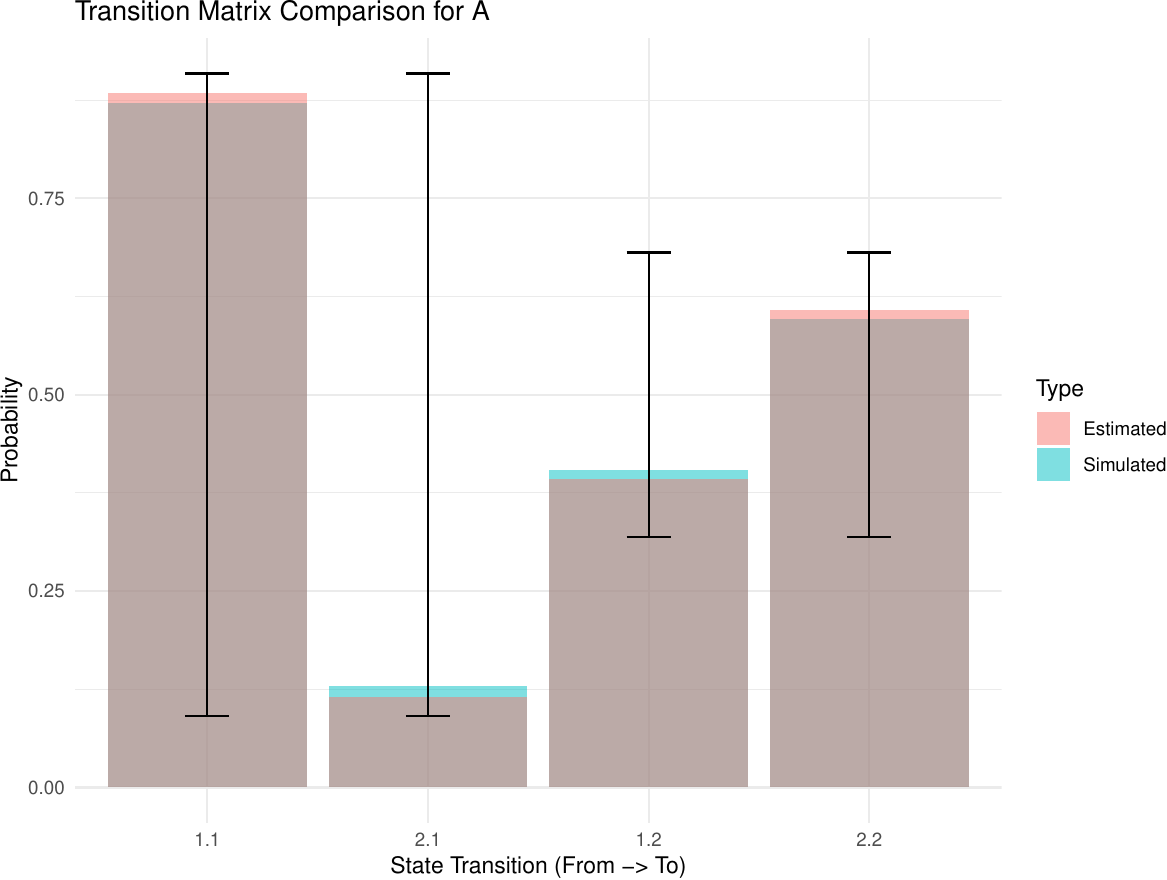}
\caption{\small Two-state transition}\label{fig:State_trans}
\end{figure}

In Figure~\ref{fig:State_trans}, we show the transition probabilities and their respective confidence intervals for the two-state model, estimated vs. simulated (with reference to the matrix in~\eqref{eq:A2}). This demonstrates the model’s capability to detect regime transitions reliably, a crucial aspect in time series analysis.

This is confirmed by the confusion matrix reported in Table~\ref{confusion_matrix} which reveals a high accuracy in classifying observations into their respective hidden states, with an overall accuracy of 97.7\%. This result is further supported by the Kappa statistic of 0.93, which reflects a strong agreement between the true and predicted state sequences. These findings confirm that the proposed model can effectively disentangle state-dependent patterns in the data while maintaining high precision.
\begin{table}[!htb]
\centering
\caption{\small Confusion Matrix two-state model. This table presents the observed states compared with the estimated states from a classification model. Rows represent the observed states, while columns indicate the predicted states.}\label{confusion_matrix}
\centering
\begin{tabular}{l|l|c|c}
\toprule
\multicolumn{2}{c|}{} & \multicolumn{2}{c}{\textbf{Estimated States}} \\
\cline{3-4}
\multicolumn{2}{c|}{} & \textbf{State 1} & \textbf{State 2} \\
\hline
\multirow{2}{*}{\textbf{Observed States}} & \textbf{State 1} & \cellcolor{gray!10}{457} & \cellcolor{gray!10}{9} \\
& \textbf{State 2} & 5 & 129 \\
\bottomrule
\end{tabular}
\end{table}


The results for the two-state model demonstrate that the FMM-HMM with skew-normal components is highly effective at capturing both symmetric and asymmetric behaviors in temporal data, while consistently identifying structural changes with reliability.

\subsection{Three-state model}

We extended the simulation study to include three hidden states in order to test the model's performance in a more complex setting. This configuration introduces additional challenges, as the increased number of regimes adds complexity to parameter estimation and state classification.
\begin{figure}[!htb]
\centering
\includegraphics[width=0.58\linewidth]{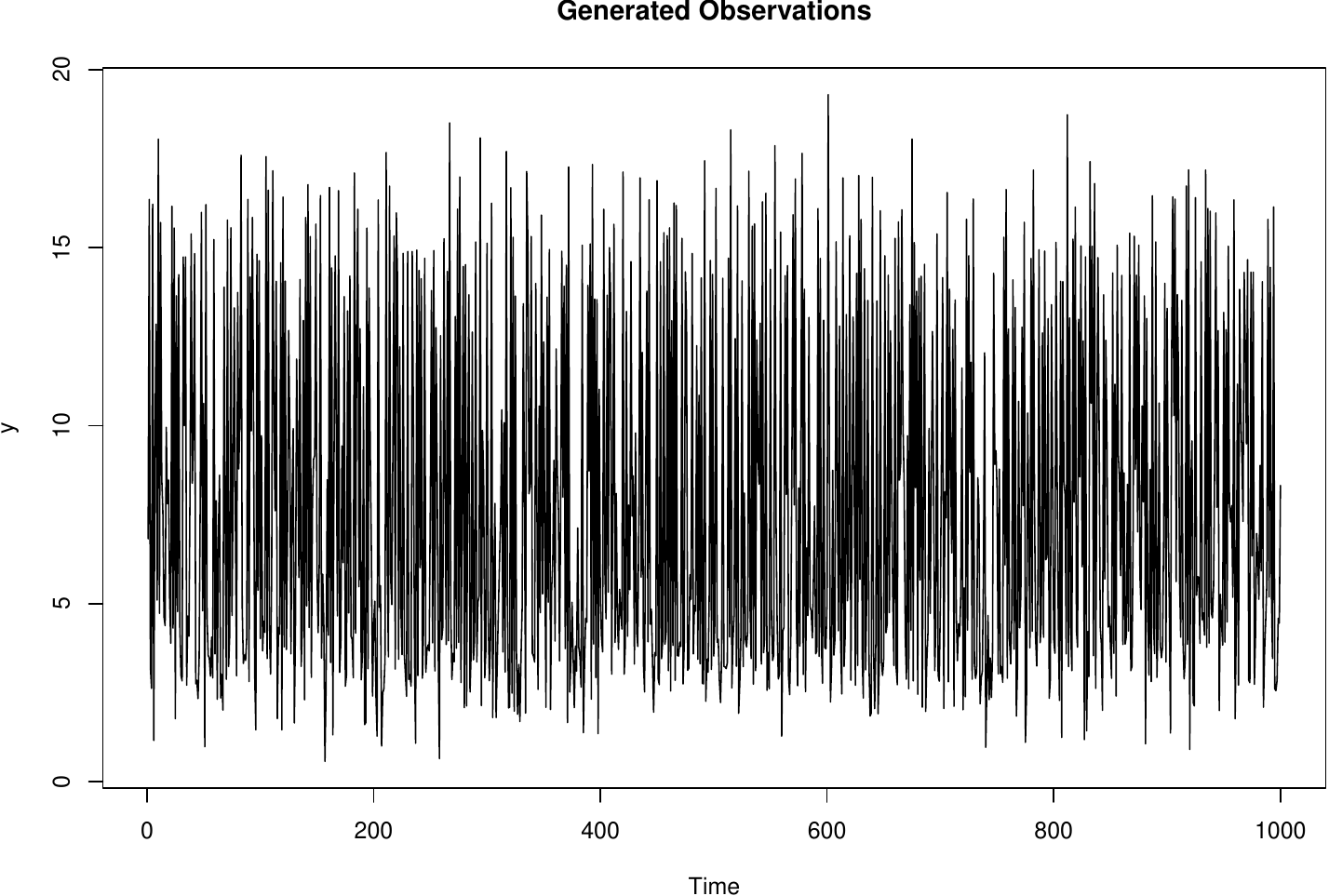}
\caption{\small Generated observations (three states)}\label{fig:Gen3}
\end{figure}

We present an additional simulation study involving the generation of 1000 data samples from a mixture of two skew-normal distributions with three states, as depicted in Figure~\ref{fig:Gen3}. The simulated parameters are summarized in Table~\ref{tab:sim2}, and the transition probabilities are simulated using the matrix in~\eqref{eq:A3}.

\begin{table}[!htb]
\centering
\caption{\small Generated Parameters for three states}
\centering
\begin{tabular}[t]{@{}lrrr@{}}
\toprule
Parameter & State 1 & State 2 & State 3 \\
\midrule
$\xi_1$ & 3.7816 & 8.3478 & 15.0385 \\
$\xi_2$ & 3.6279 & 9.9347 & 14.7152 \\
\midrule
$\omega^2_1$ & 0.1237 & 0.8513 & 1.1937 \\
$\omega^2_2$ & 1.1986 & 0.4675 & 1.7938 \\
\midrule
$\lambda_{1}$ & 0.5110 & 0.5396 & -0.1079 \\
$\lambda_{2}$ & 0.0172 & -0.2152 & 0.4528 \\
\midrule
$\zeta$ & 0.0050 & 0.8715 & 0.1234 \\
\bottomrule
\end{tabular}\label{tab:sim2}
\end{table}



\paragraph{Results of the three-state model}





\begin{figure}[!htb]
\centering
\includegraphics[width=0.46\linewidth]{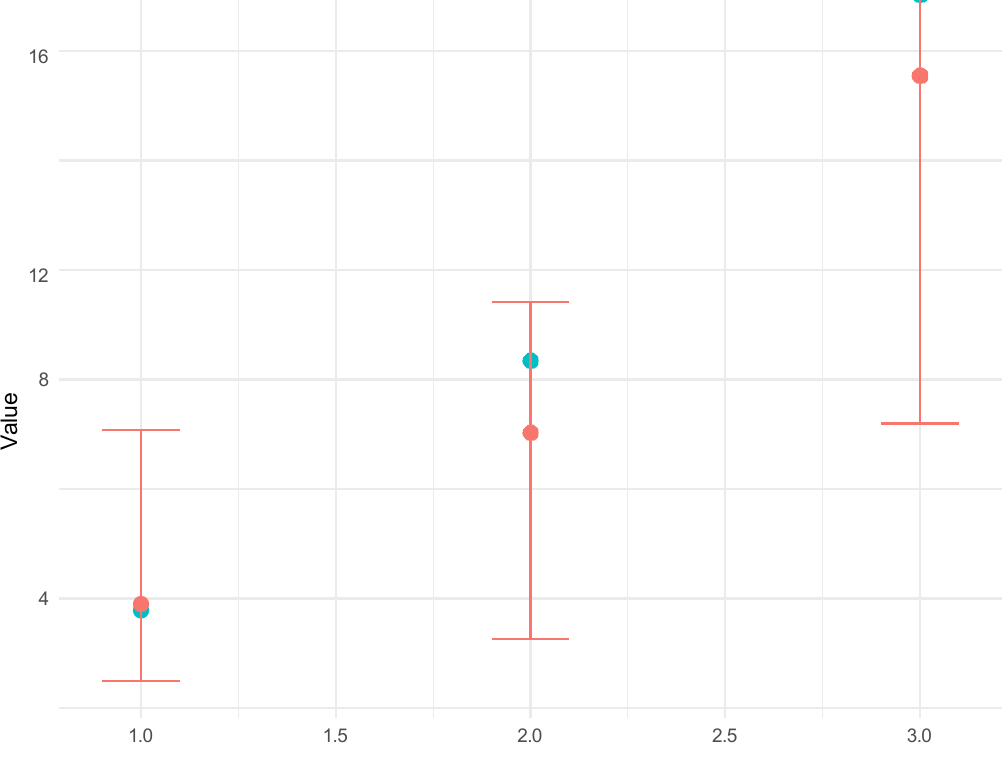}
\includegraphics[width=0.46\linewidth]{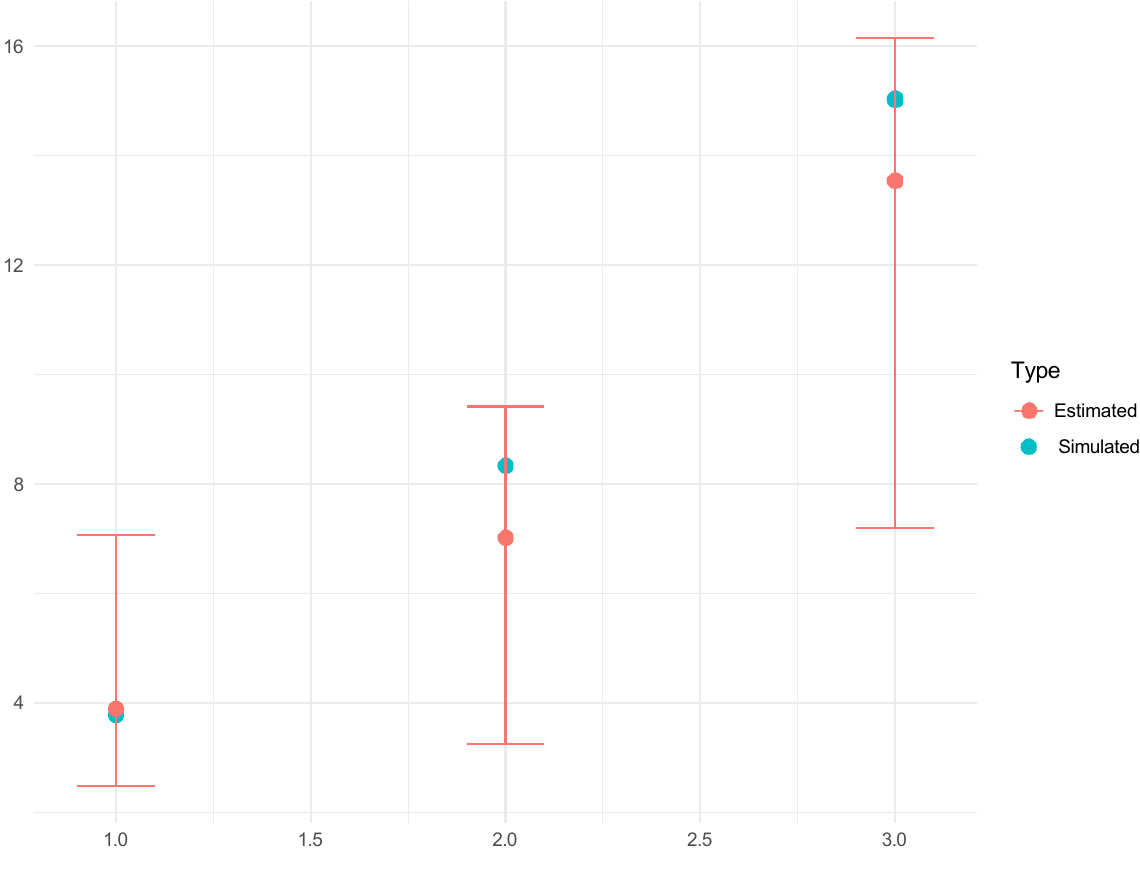}
\par\vspace{0.5cm} 

\includegraphics[width=0.46\linewidth]{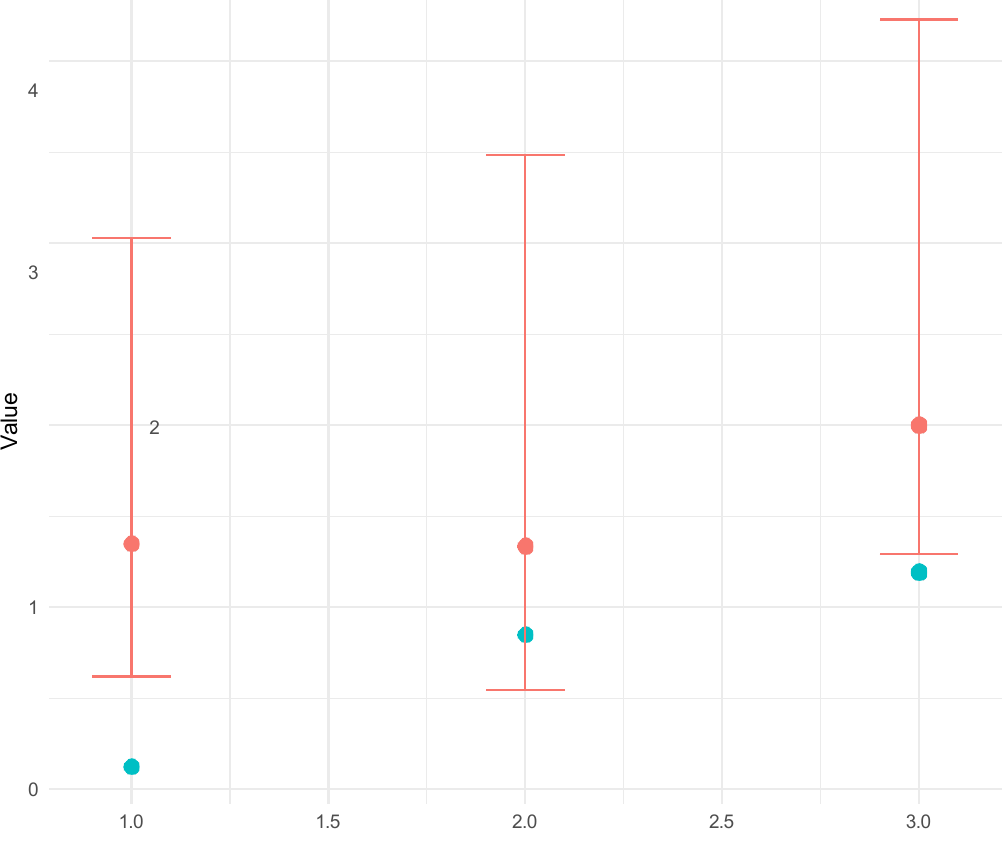}
\includegraphics[width=0.46\linewidth]{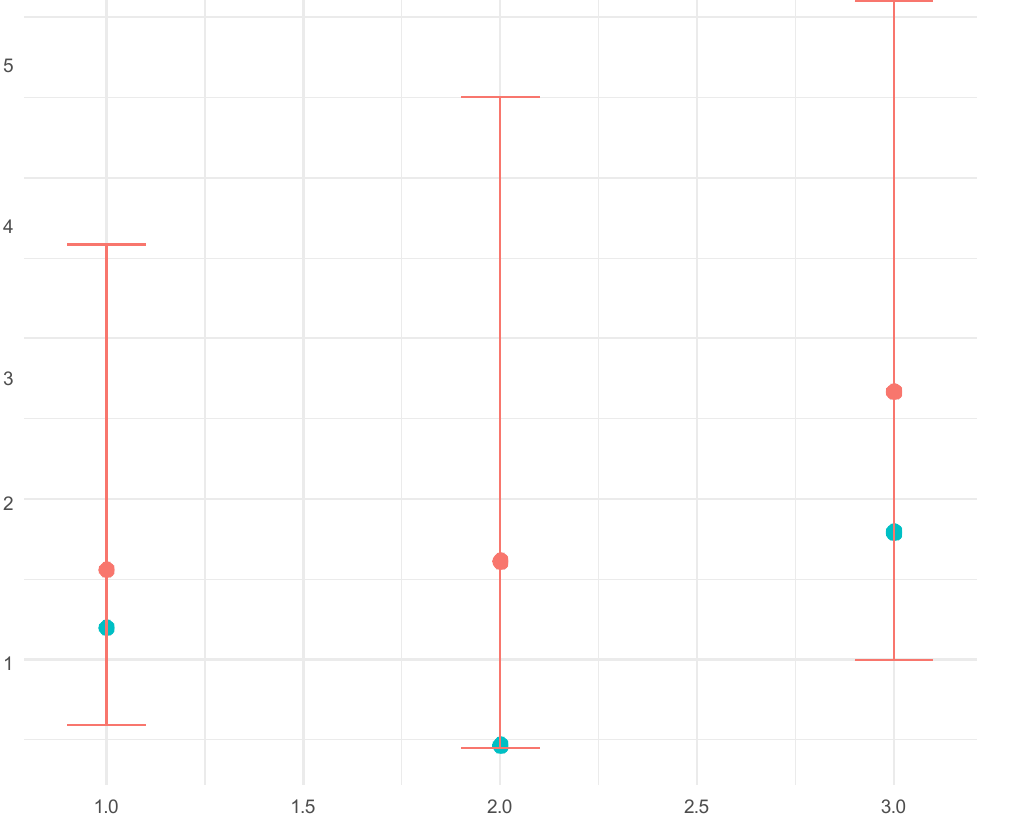}
\par\vspace{0.5cm} 

\includegraphics[width=0.46\linewidth]{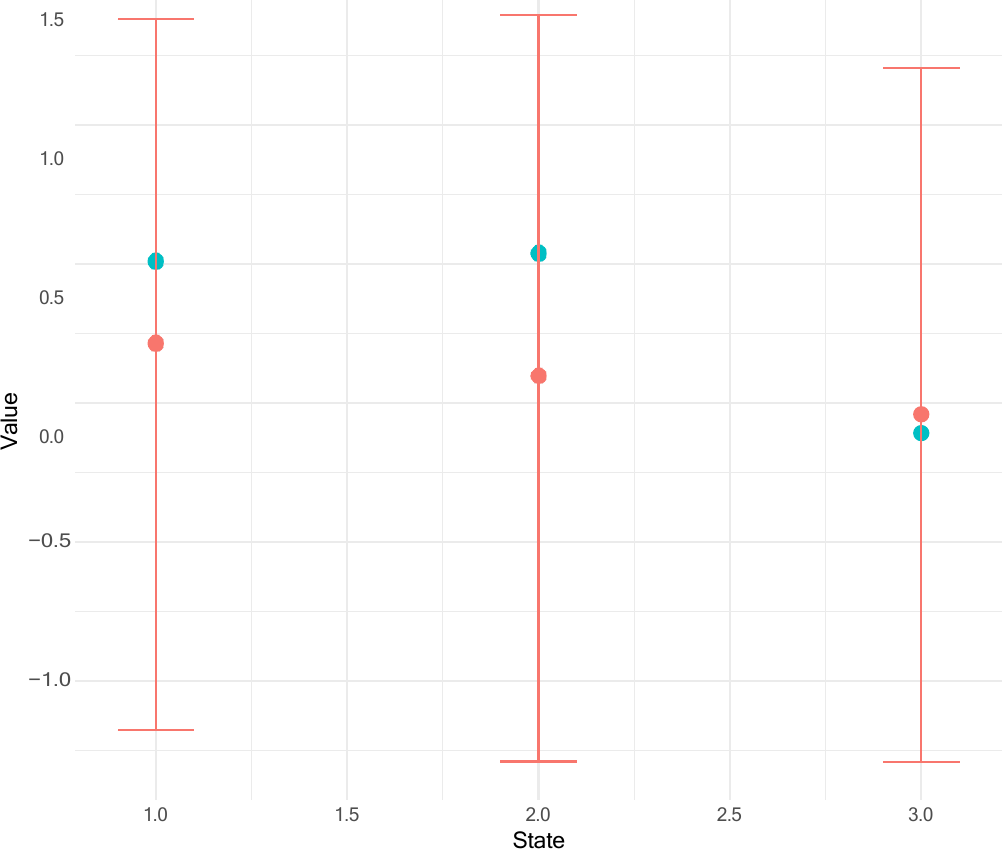}
\includegraphics[width=0.46\linewidth]{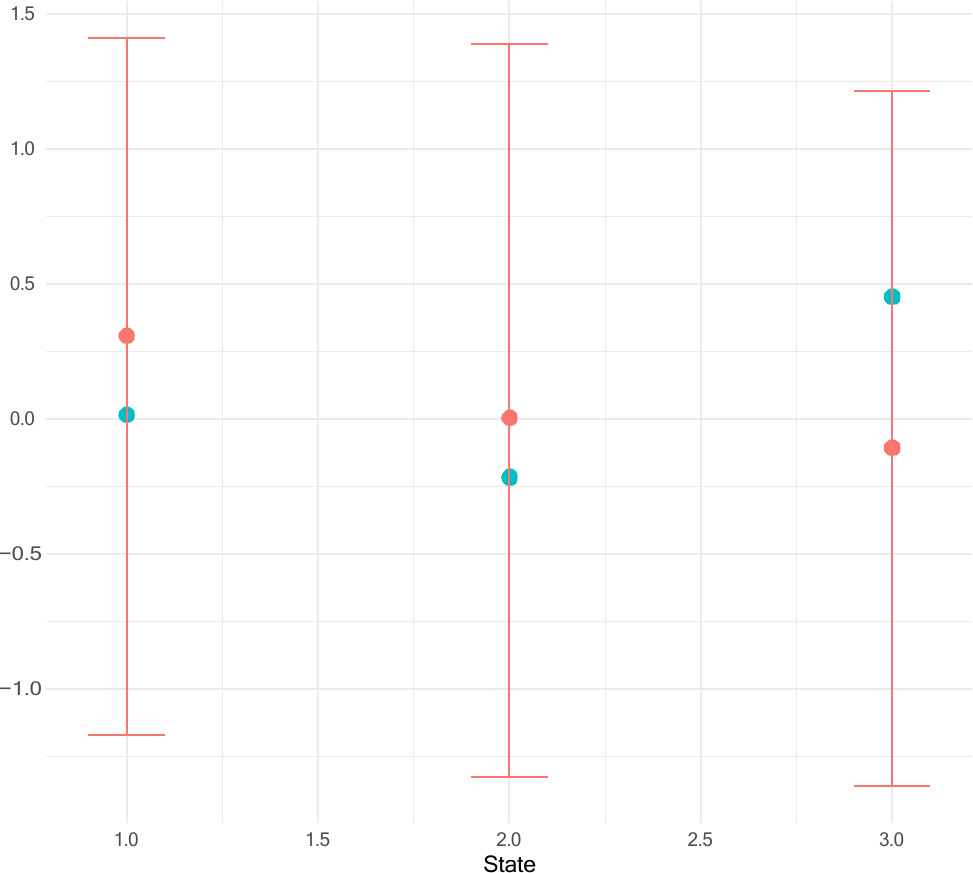}
\caption{\small Comparison between simulated and estimated values for location, scale, and skewness. Top row: location values. Middle row: scale values. Bottom row: skewness values.}
\label{fig:comparison3}
\end{figure}

The values of the simulated parameter are reported in Table~\ref{tab:sim2}, while we can see here how the model effectively recovers the location and scale parameters in the three hidden states. The graphic representation may be traced in Figure~\ref{fig:comparison3}.





However, the skewness parameters exhibit slightly greater variability than in the two-state case; this can be attributed to weaker asymmetry signals in certain regimes (see Figure~\ref{fig:comparison2}). However, the overall accuracy in parameter estimation demonstrates the robustness of the model, even with increased complexity. 

\begin{figure}[!htb]
\centering
\includegraphics[width=0.7\linewidth]{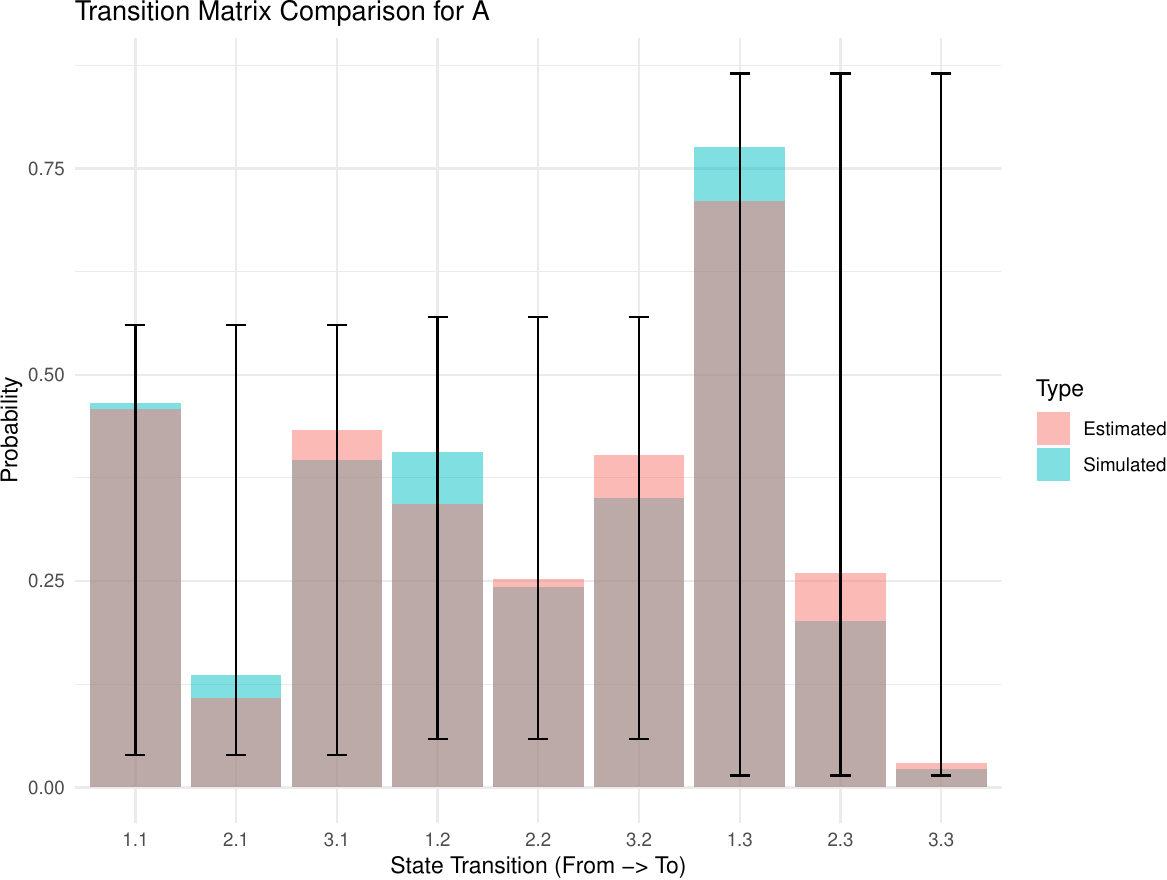}
\caption{\small Three-state transition}\label{fig:three_state_matrix}
\end{figure}

The estimated transition probability matrix for the three-state model is presented in Figure~\ref{fig:three_state_matrix}. The comparison between estimated and simulated transitions is illustrated along with their respective confidence intervals. In this case, the proposed model accurately captures the state-switching behavior, with probabilities that once again closely match the simulated values from matrix \textbf{A} in~\eqref{eq:A3}. This result highlights the model’s ability to handle more intricate transition structures, ensuring reliable identification of dynamic regime changes.
\begin{equation}
\mathbf{A}=\left(\begin{array}{cccc}
0.4661 & 0.1368 & 0.3972 \\
0.4069 & 0.2424 & 0.3507 \\
0.7762 & 0.2020 & 0.0218 \\
\end{array}\right)\label{eq:A3}
\end{equation}

The confusion matrix for the three-state model, presented in Table~\ref{confusion_matrix3}, reveals an overall accuracy of 96.1\%. Relative statistics further emphasize a Kappa value of 0.93. This high level of accuracy demonstrates that the model retains a strong classification performance, even with the increased complexity introduced by the third state. The model’s ability to accurately assign observations to their respective regimes highlights its robustness and potential for addressing real-world problems involving multiple structural changes.
\begin{table}[!htb]
\centering
\caption{\small Confusion Matrix three-state model. The observed states are compared with the estimated states from a classification model. Rows represent the observed states, while columns indicate the predicted states.}
\centering
\begin{tabular}{@{}l|l|c|c|c@{}}
\toprule
\multicolumn{2}{c|}{} & \multicolumn{3}{c}{\textbf{Estimated States}} \\
\cline{3-5}
\multicolumn{2}{c|}{} & \textbf{State 1} & \textbf{State 2} & \textbf{State 3} \\
\hline
\multirow{3}{*}{\textbf{Observed States}} & \textbf{State 1} & 554 & 2 & 0 \\
& \textbf{State 2} & 14 & 114 & 22 \\
& \textbf{State 3} & 0 & 1 & 293 \\
\bottomrule
\end{tabular}
\label{confusion_matrix3}
\end{table}


\section{Gender gap analysis in human mortality}\label{sec:realdata}

We consider historical data sourced from \cite{HMD}, which provides standardized mortality rates and life tables covering both sexes in various age groups and periods. Based on the study by \cite{bufaloNigri2024}, we utilize the gender gap variable, in which a mixture of two skew-normal distributions can approximate the temporal distribution. The gender gap variable is the logarithm of the ratio between the number of deaths and the exposed population.

\begin{figure}[!htb]
\centering
\includegraphics[width=0.8\linewidth]{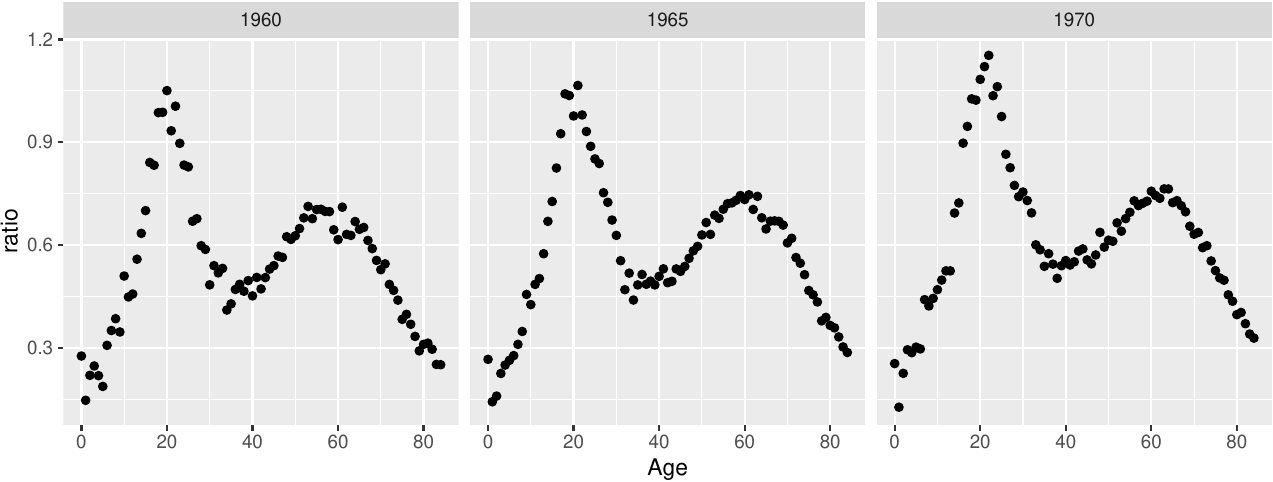}\\
\includegraphics[width=0.8\linewidth]{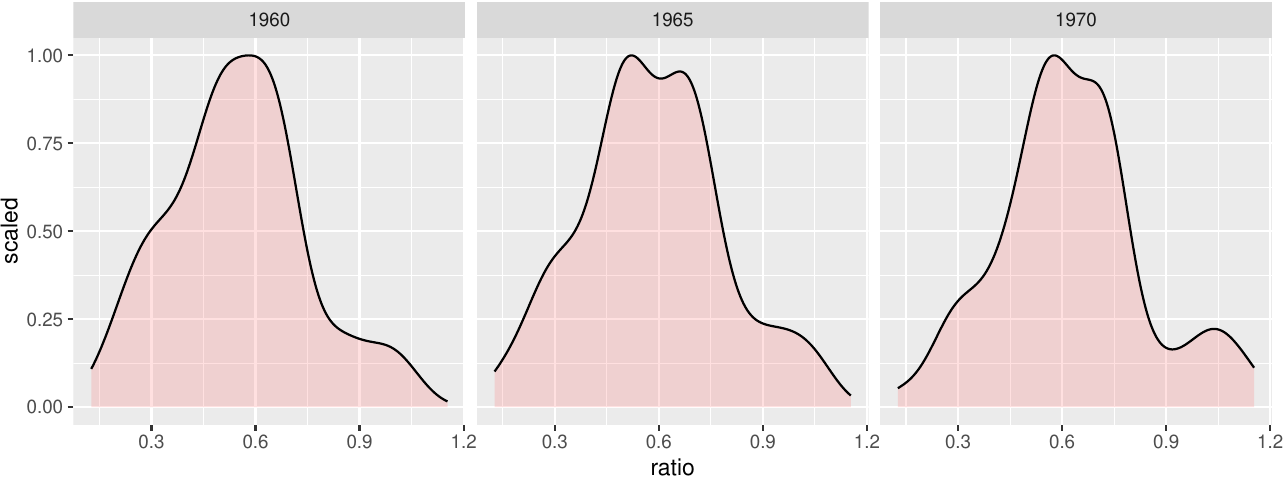}
\par\vspace{0.5cm} 
\caption{\small Empirical distribution of gender gap in mortality. Data are sourced from the \cite{HMD}.}
\label{fig:realdata}
\end{figure}

Using data from 0 to 90 years for the period 1960-1975, the mortality rate is calculated from death counts $D^{g}_{x,t}$ and population exposure $ E^g_{x,t}$ as follows:
\begin{equation*}
m^{g}_{x,t} = \frac{D^g_{x,t}}{E^g_{x,t}},
\end{equation*} 
where \( m^g_{x,t} \) represents the age-specific mortality rate for a given sex \( g \), age $x$, and year \( t \). Additionally, the gender mortality disparity is quantified using the log-transformed mortality rate ratio:
\begin{equation}
GP_{x,t} = \log \left(\frac{m_{x,t}^M}{m_{x,t}^F} \right).
\end{equation}
In this study, we focus on mortality data from the United States (US), as provided by \cite{HMD}; a graphical representation is provided in Figure \ref{fig:realdata}. 

\subsection{Two-state and three-state modelings}\label{sec:4.1}

\begin{figure}[!htb]
\centering
\includegraphics[width=0.48\linewidth]{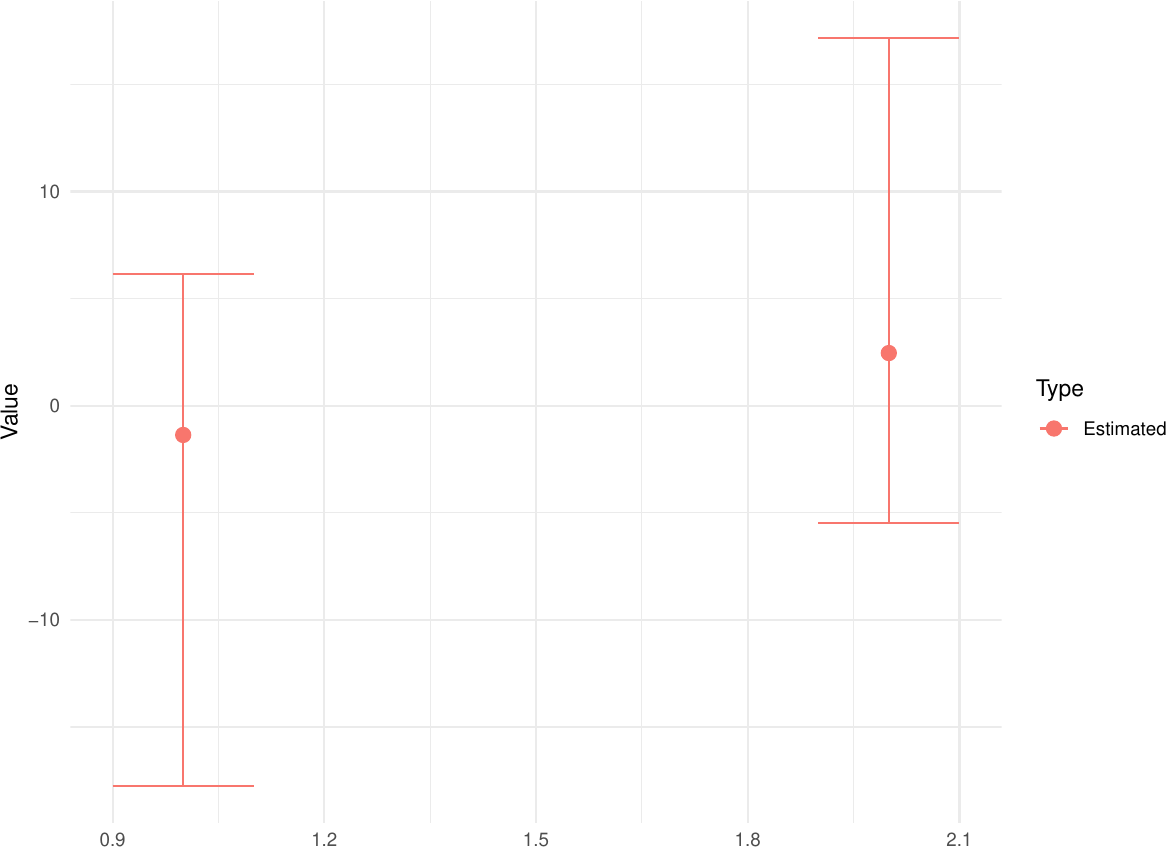}
\includegraphics[width=0.48\linewidth]{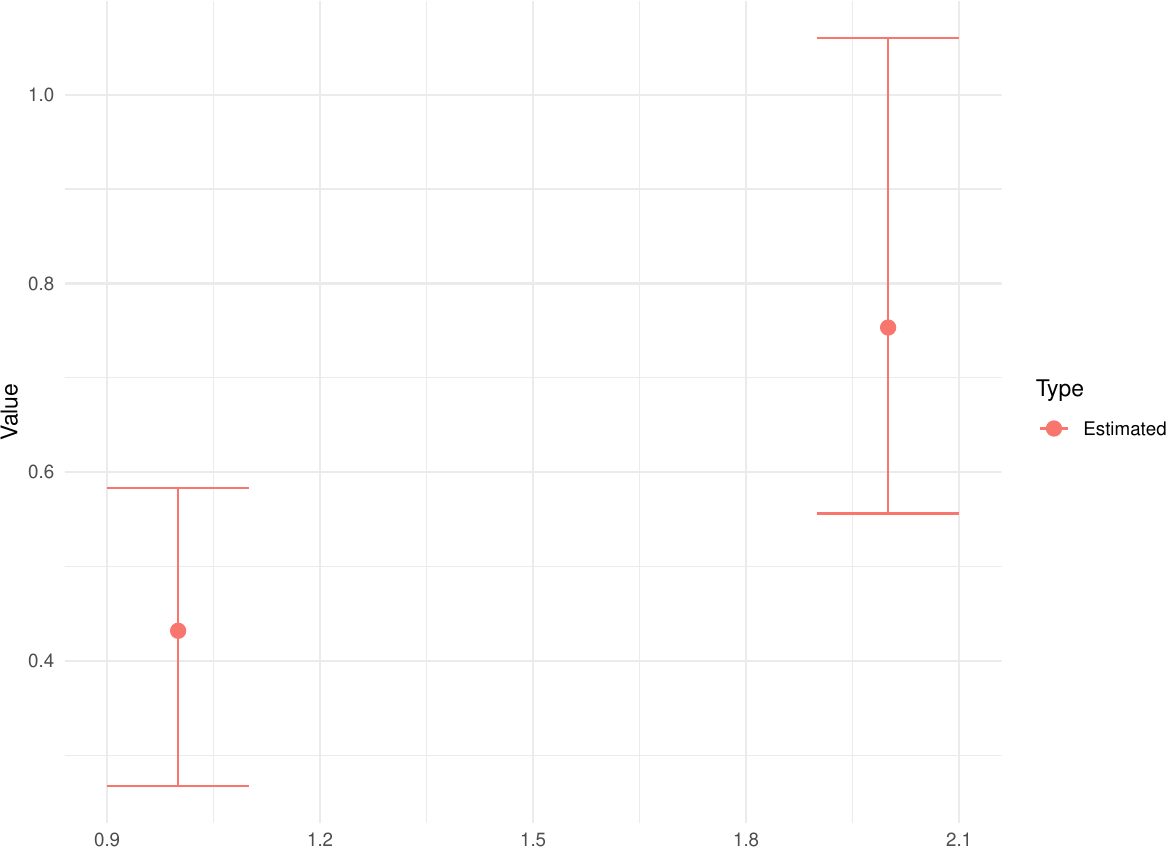}
\par\vspace{0.5cm} 
\includegraphics[width=0.48\linewidth]{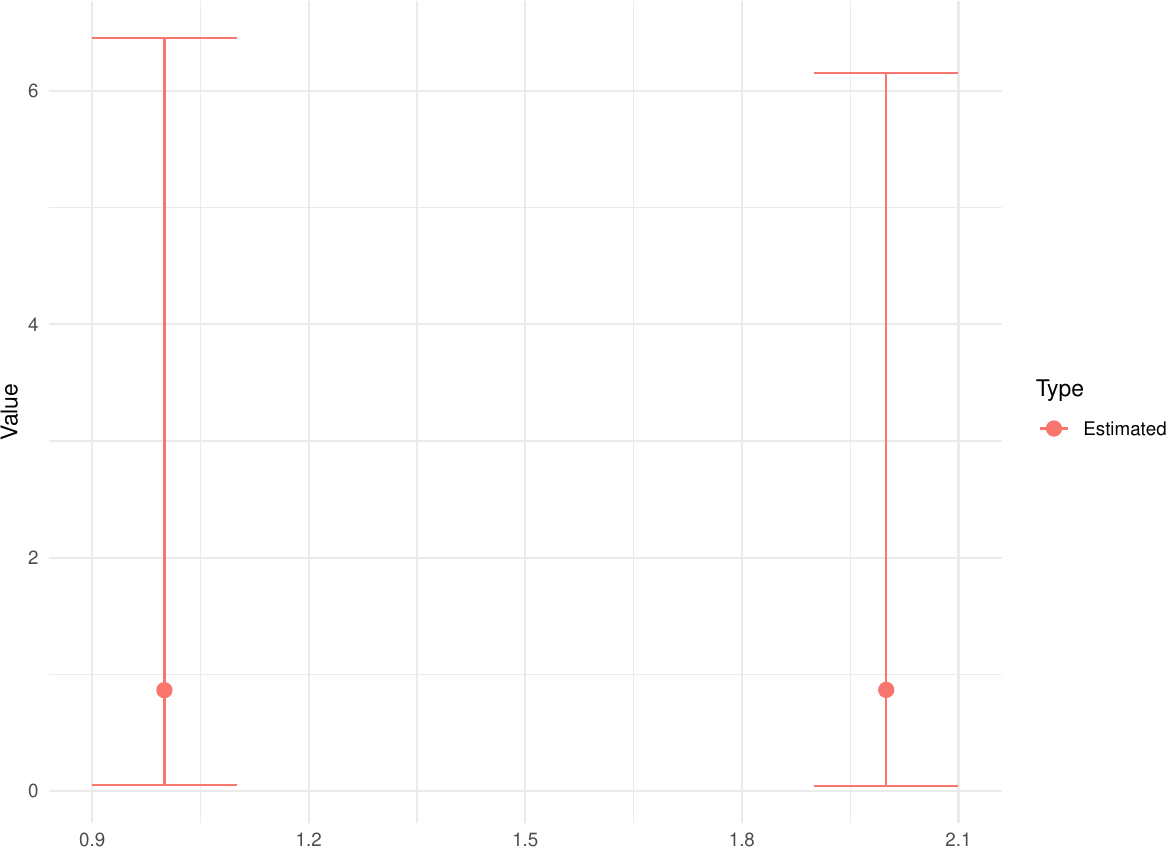}
\includegraphics[width=0.48\linewidth]{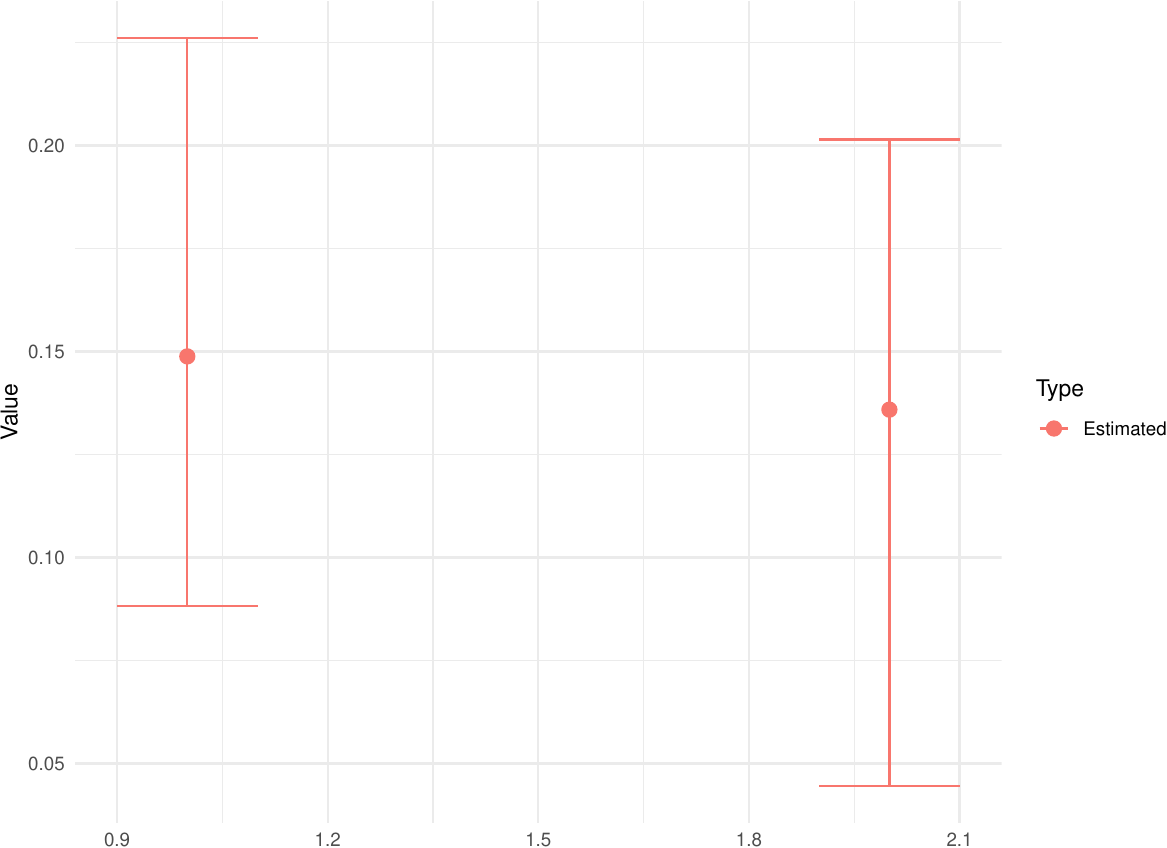}
\par\vspace{0.5cm} 
\includegraphics[width=0.48\linewidth]{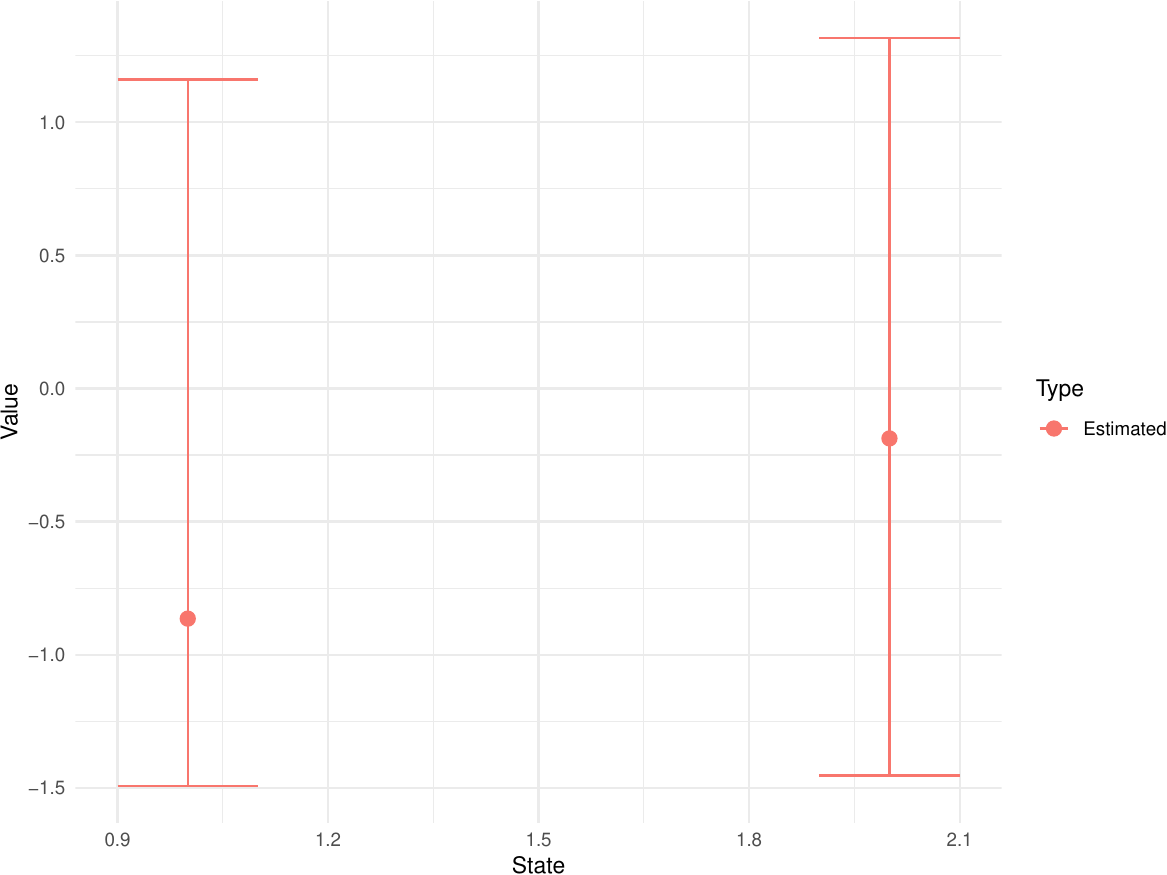}
\includegraphics[width=0.48\linewidth]{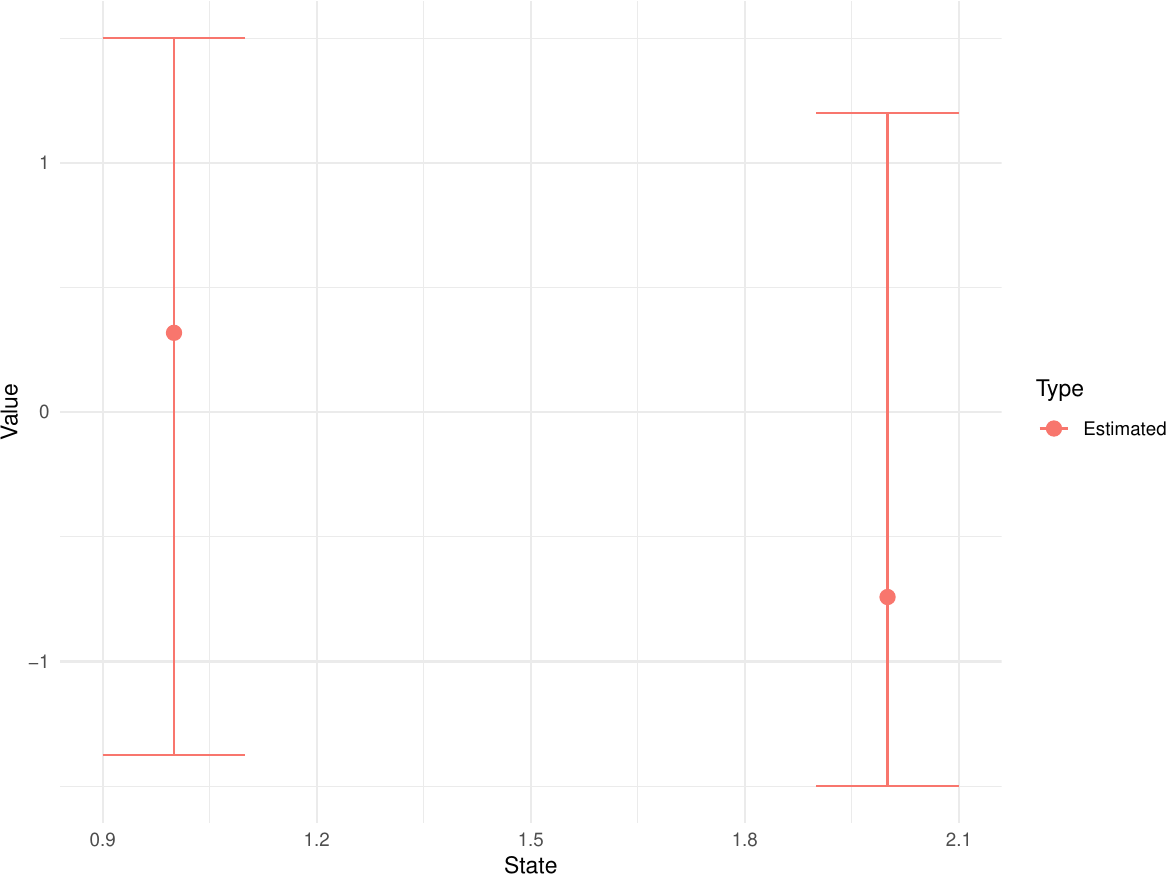}
\caption{\small Estimated values for location, scale, and skewness on gender gap mortality data using two-state modeling. Top row: location values. Middle row: scale values. Bottom row: skewness values.}
\label{fig:estimate_realdata2}
\end{figure}

\begin{figure}[!htb]
\centering
\includegraphics[width=0.495\linewidth]{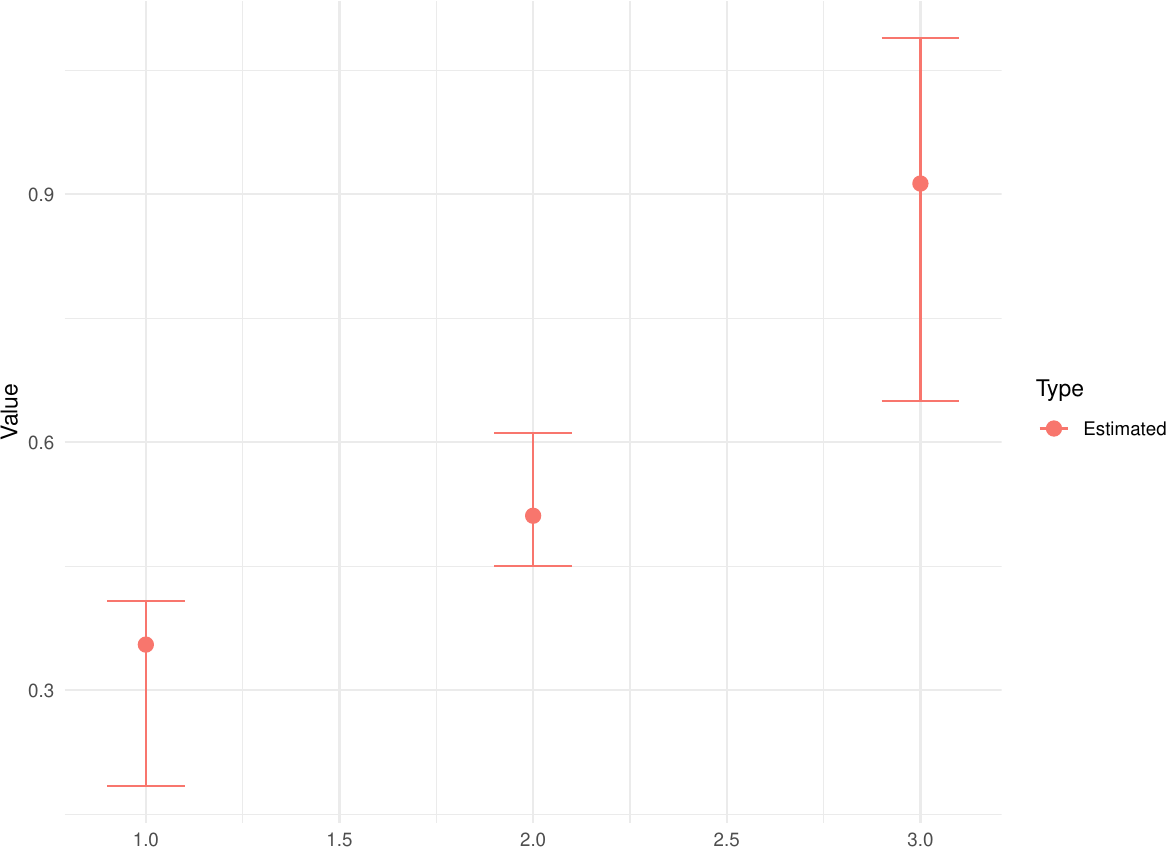}
\includegraphics[width=0.495\linewidth]{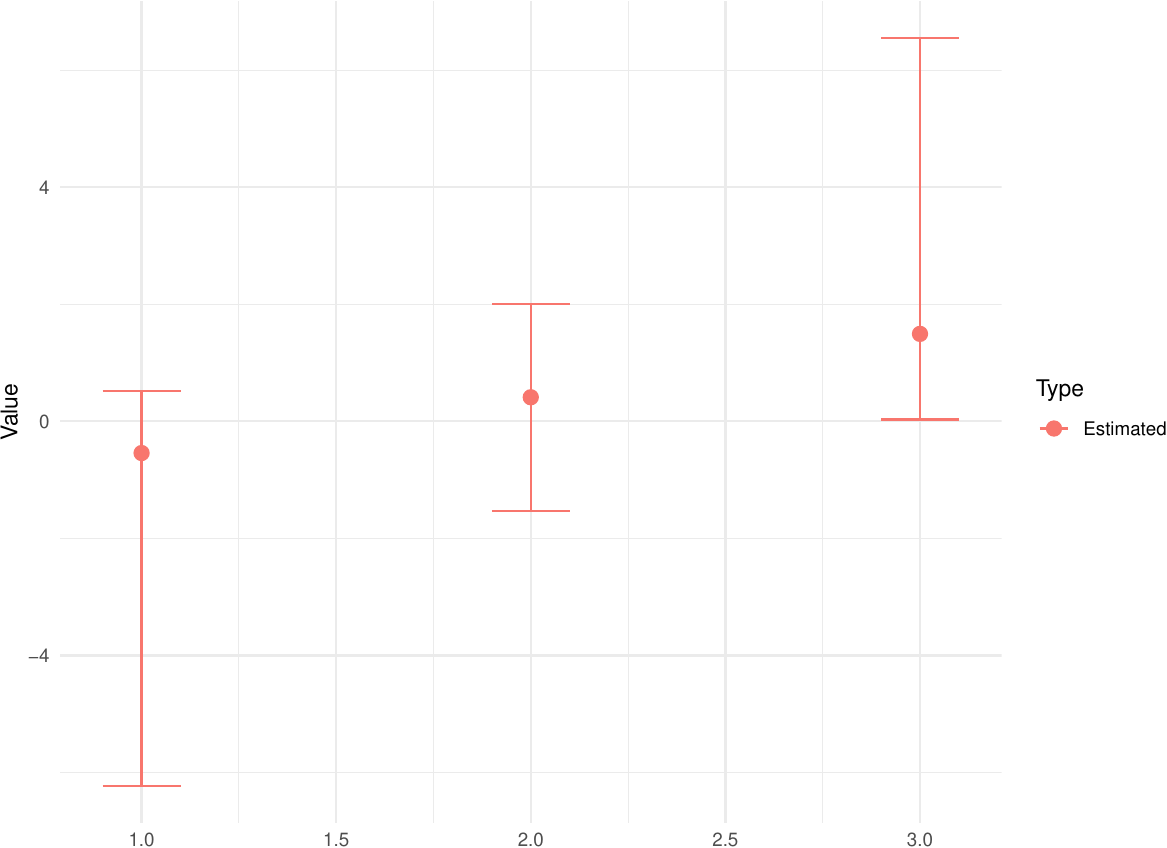}
\par\vspace{0.5cm}

\includegraphics[width=0.495\linewidth]{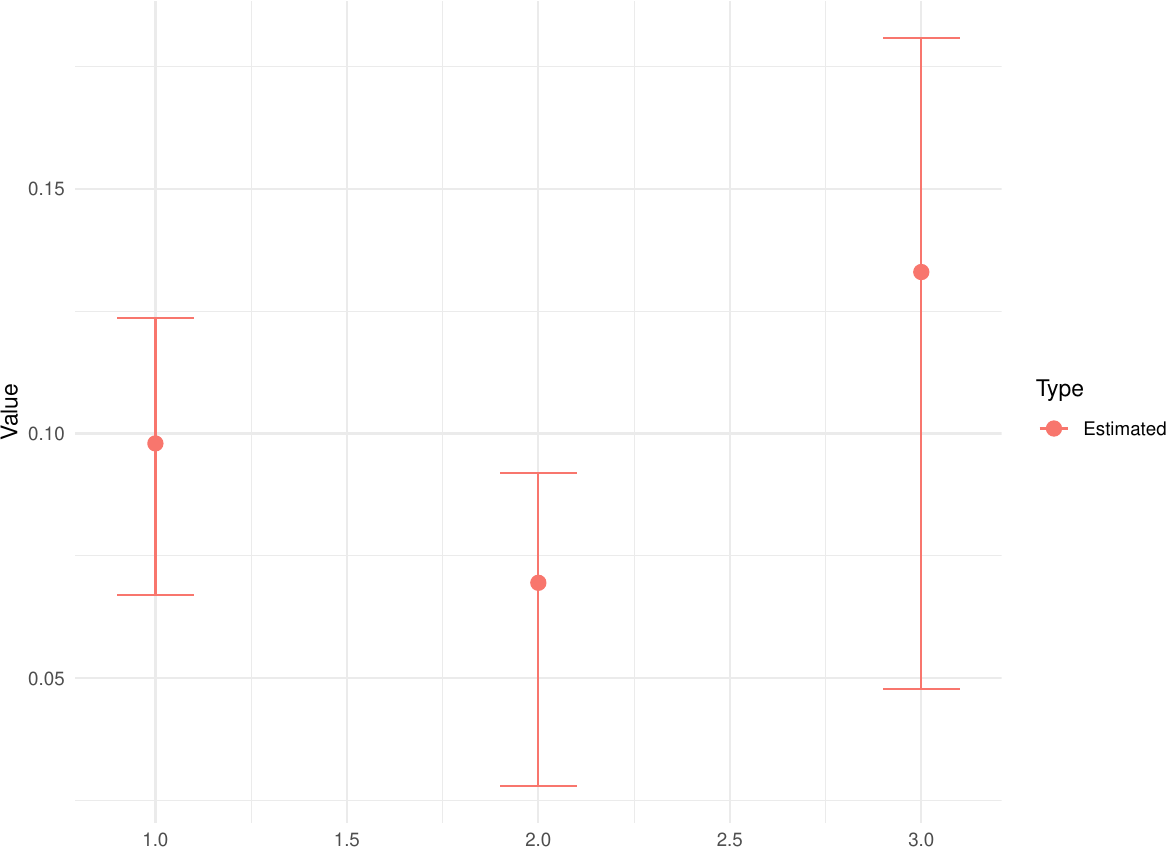}
\includegraphics[width=0.495\linewidth]{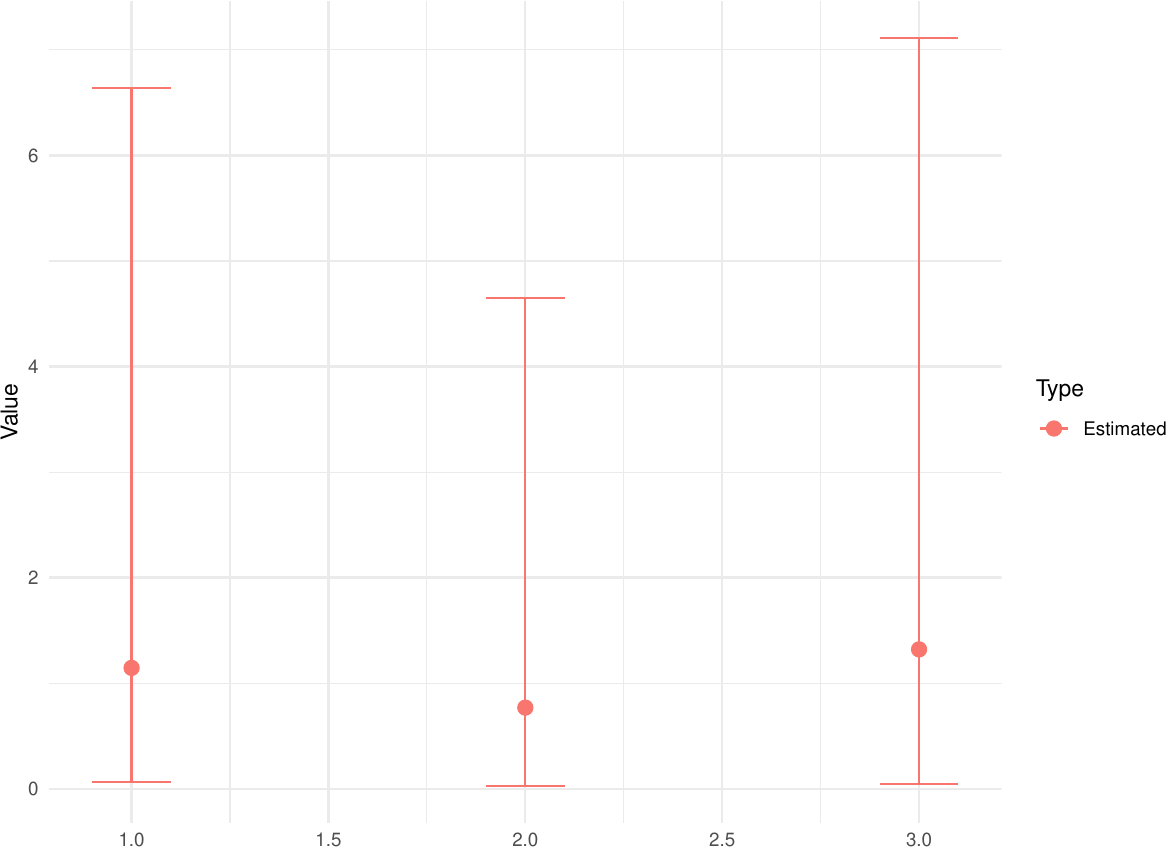}
\par\vspace{0.5cm} 

\includegraphics[width=0.495\linewidth]{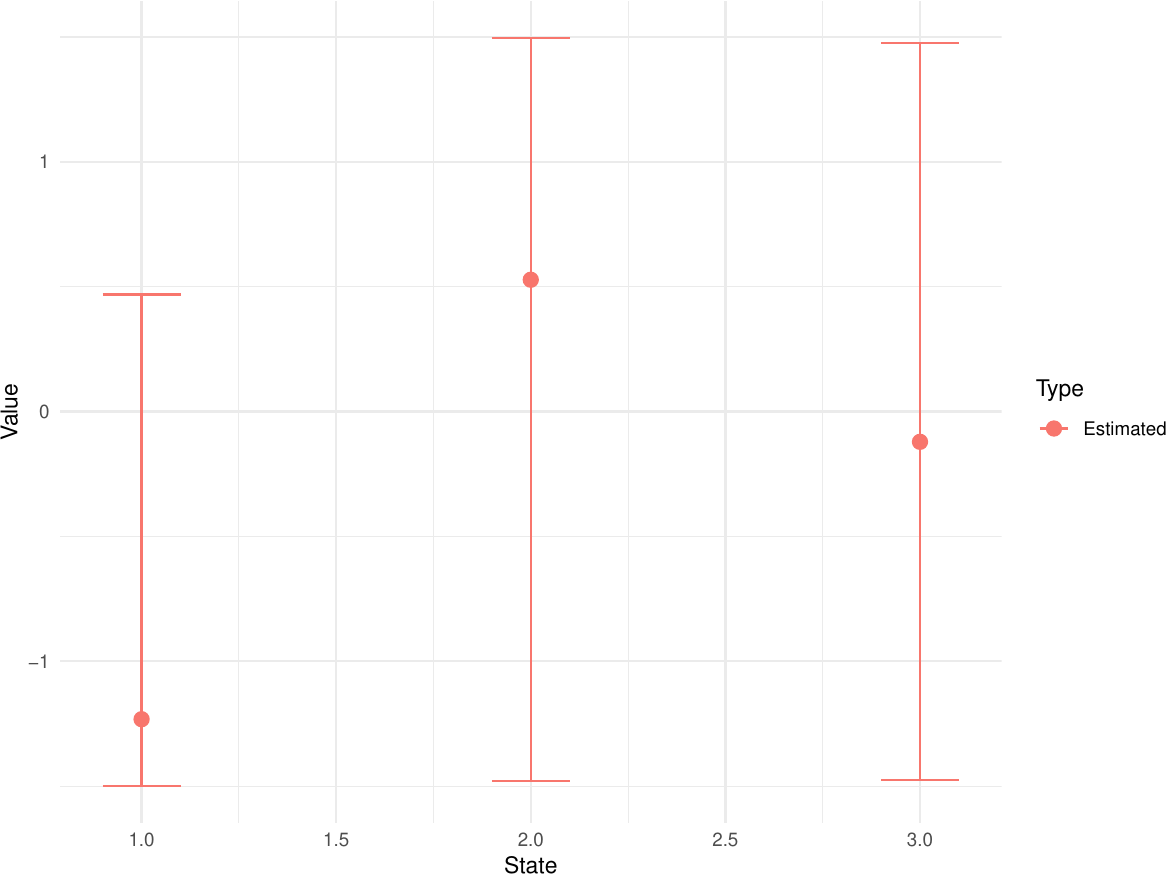}
\includegraphics[width=0.495\linewidth]{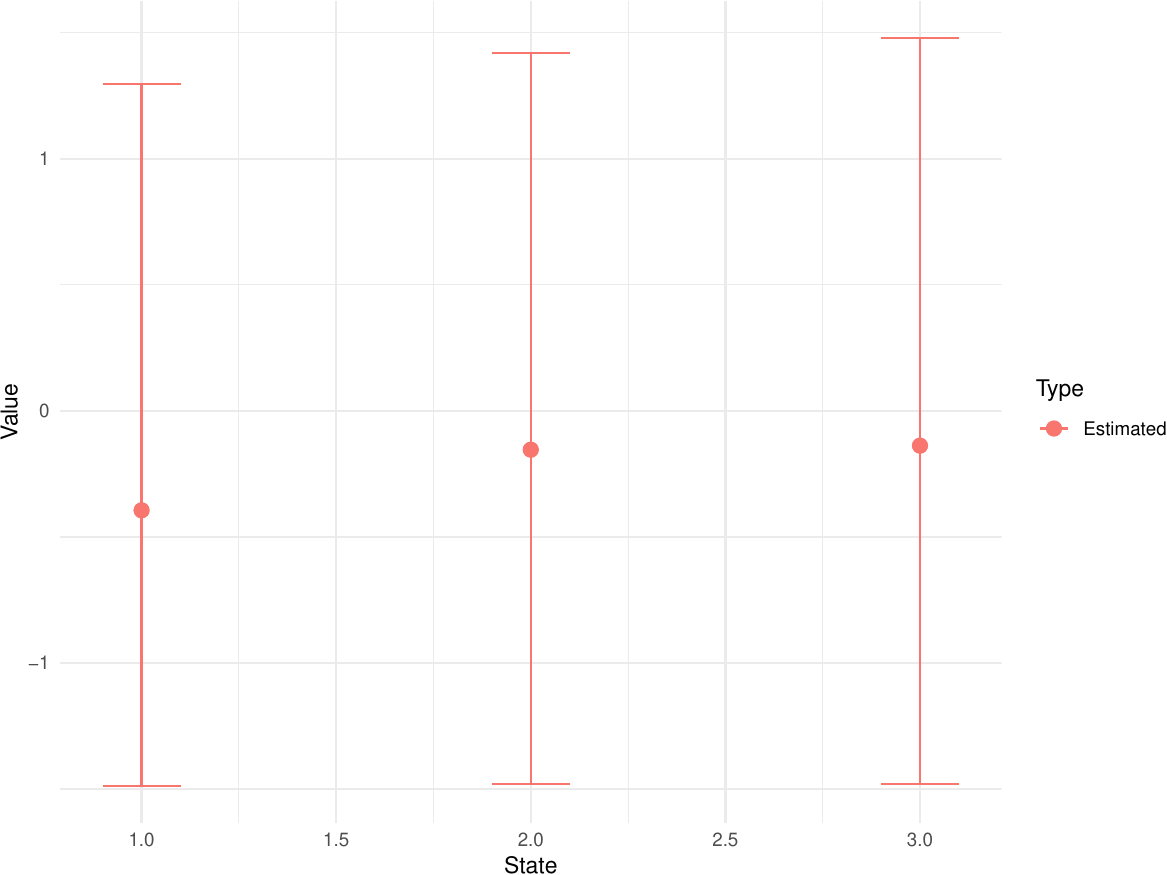}
\par\vspace{0.5cm} 
\caption{\small Estimated values for location, scale, and skewness on gender gap mortality data using three-state modeling. Top row: location values. Middle row: scale values. Bottom row: skewness values.}
\label{fig:estimate_realdata3}
\end{figure}

We compare the estimates from the model using two and three states, respectively. The graphical output is provided in Figure~\ref{fig:estimate_realdata2} for the two-state model and in Figure~\ref{fig:estimate_realdata3} for the three-state model.

Their respective transition probability matrices~\eqref{eq:A2_real_data} and~\eqref{eq:A3_real_data}.
\begin{equation}
\mathbf{A}=\left(\begin{array}{cccc}
0.954 & 0.0453 \\
0.0513 & 0.9486 \\
\end{array}\right)\label{eq:A2_real_data}
\end{equation}

\begin{equation}
\mathbf{A}=\left(\begin{array}{cccc}
0.9425 & 0.0537 & 0.0036 \\
0.0360 & 0.9247 & 0.0392 \\
0.0039 & 0.0775 & 0.9185 \\
\end{array}\right)\label{eq:A3_real_data}
\end{equation}

The models are also compared using the Bayesian Information Criterion (BIC), which is defined as:
\begin{equation*}
\mathrm{BIC} = -2 \cdot \log \mathcal{L}(\widehat{\theta}) + p \cdot \log(N),
\end{equation*}
where $\log$ denotes the natural logarithm, $\mathcal{L}(\widehat{\theta})$ is the maximum likelihood of the model, $p$ is the total number of estimated parameters and $N$ is the number of observations. The information criterion suggests that the solution associated with the lowest value should be preferred, as it offers the best trade-off between goodness of fit and model complexity. In our case study, this corresponds to $100.11$ for the two-state model and $171.62$ for the three-state model; thus, empirical evidence suggests that the two-state model is the most appropriate to describe the phenomenon under study.

Such results are further confirmed by the integrated complete likelihood (ICL) \citep{bertoletti2015ICL, Cleynen2014ICL}, with a value of $-264.3358$ and $-243.4085$ for the two- and three-state respectively. The following formula mathematically defines ICL:
\begin{equation*}
\text{ICL}(K) = \text{BIC}(K) - \mathcal{H}(\mathbf{Z}).
\end{equation*}

Practically, the ICL adds an entropy term $\mathcal{H}(\mathbf{Z})$ to the BIC, which is defined as:
\begin{equation*}
\mathcal{H}(\mathbf{Z}) = - \sum_{t=1}^{n} \sum_{k=1}^{K} \gamma_{tk} \log \gamma_{tk},
\end{equation*}
where $\gamma_{tk}$ represents the probability that observation $t$ belongs to the cluster $k$.

In essence, to highlight the difference between BIC and ICL, we can say that while the former evaluates only the marginal likelihood (integrated over the parameters) of the observed data, the latter also accounts for the latent assignments, thereby penalizing uncertainty in clustering. The ICL is therefore more conservative than the BIC, tending to select a smaller number of clusters, as it favors models that not only fit the data well but also produce well-separated and less ambiguous latent assignments. 

To provide a substantive interpretation of the estimated states, we contextualize the regimes within the demographic and epidemiological landscape of the United States in the 1960–1975 period. 

Specifically, \textbf{State 1} corresponds to years characterized by a \emph{larger gender gap} in mortality rates, which can be attributed to persistent behavioral and occupational risk factors that disproportionately affect male mortality. These include a higher prevalence of smoking among men, more dangerous work environments, and delayed response to emerging public health interventions. Furthermore, during this period, male cohorts still experienced the long-term health consequences of previous life exposures.

Conversely, \textbf{State 2} reflects a \emph{narrower gender gap} and can be associated with initial signs of improvement in male mortality and greater parity in health outcomes. This shift may be driven by early anti-smoking campaigns, broader access to healthcare, and demographic changes such as declining fertility and improvements in maternal and child health, which indirectly influence adult female mortality.

Although the three-state model introduces an additional regime, its interpretability is less straightforward. We conjecture that \textbf{State 3} may represent \emph{transitional periods} with mixed dynamics or cohort effects not fully captured by the other two states. Because of the higher BIC and reduced parsimony, we regard the two-state model as the more reliable representation of the structural patterns in the data.

\subsection{Sensitivity analysis with respect to prior distributions}

To assess the robustness of our results to the choice of prior distributions, we performed a sensitivity analysis in Table~\ref{tab:priorsensitivity} considering different combinations of priors for the main parameters of the model. In addition to the baseline configuration --- which adopts priors for location, scale, and skewness parameters (Normal(0, 10), Cauchy(0, 2), and Normal(0, 1), respectively) and uniform Dirichlet priors for the transition probabilities and mixture weights - we also explored other prior alternatives. Specifically, we tested priors with decreased variance (Normal(0,5), Normal(0,0.5)), and priors centered on the empirical mean of the observed data. We also varied the parameters of the Dirichlet distribution to assess the effect of favoring more frequent or rarer transitions between states \textit{a priori}.


\begin{table}[!htb]
\centering
\tabcolsep 0.33in
\caption{\small A combination of different prior choices.}\label{tab:priorsensitivity}
\begin{tabular}{@{}lcccc@{}}
\toprule
Scenario & $\mu_1, \mu_2$ & $\sigma_1, \sigma_2$ & $\text{skew}_1, \text{skew}_2$ & $A$, $\theta$ \\
\midrule
Baseline & $\mathcal{N}(0,10)$ & $\mathrm{Cauchy}(0,2)$ & $\mathcal{N}(0,1)$ & Dirichlet(1) \\
S1   & $\mathcal{N}(0,5)$  & $\mathcal{N}(0,2)\;T[0.01,10]$ & $\mathcal{N}(0,0.5)$ & Dirichlet(1) \\
S2   & $\mathcal{N}(0,1)$ & $\mathrm{Cauchy}(0,1)$ & $\mathcal{N}(0,0.5)$ & Dirichlet(1) \\
\bottomrule
\end{tabular}
\end{table}

In the case of the real data study that we outlined at the beginning of this section, the BIC effectively identifies the number of states, but it fails to discriminate between scenarios characterized by different priors, suggesting a model that is essentially insensitive to their choice. However, in the contexts involving Hidden Markov Models -- or more generally, mixture models -- model complexity may not be adequately penalized in the presence of overlapping components.

Table~\ref{tab:bic_icl} provides a general overview of the results corresponding to the scenarios presented in Table~\ref{tab:priorsensitivity}, with a special emphasis on the baseline formulation.
\begin{table}[!htb]
\centering
\tabcolsep 0.22in
\caption{\small BIC and ICL values for different models and state configurations}\label{tab:bic_icl}
\begin{tabular}{@{}lcccccc@{}}
\toprule
& \multicolumn{2}{c}{Baseline} & \multicolumn{2}{c}{Scenario~1} & \multicolumn{2}{c}{Scenario~2} \\
\cmidrule(lr){2-3}\cmidrule(lr){4-5}\cmidrule(lr){6-7}
& 2 states & 3 states & 2 states & 3 states & 2 states & 3 states  \\
\midrule
BIC & 100.1098 & 171.6168 & 100.1098 & 171.6168 & 100.1098 & 171.6168 \\
ICL & -264.3358 & -243.4085 & -112.4876 & -92.2047 & -53.4818 & -58.2559 \\
\bottomrule
\end{tabular}
\end{table}

Based on this, we can confirm the choice of the baseline model in all scenarios analyzed, highlighting both the robustness of the results and a low sensitivity to the choice of priors. Across all scenarios considered, the estimates of the main quantities of interest, in particular, the assignment to hidden states and the estimation of component parameters, remained stable with respect to the choice of priors, confirming that inference is primarily driven by the information contained in the data. More noticeable differences only emerged in the presence of weakly informative data or for components that were poorly represented, a well-known situation in the literature on skew-normal mixture models. These findings are consistent with the methodological recommendations in \cite{yin2020variable}, \cite{zeng2023variable}, and \cite{chen2025}.






\section{Conclusion}\label{sec:conc}

We propose a novel framework that combines a mixture model based on the skew-normal distribution with an HMM. This approach assumes a constant transition probability matrix over time while allowing the emission distributions to evolve according to the hidden states (regimes). This design enables the exploration of dynamic dependencies between variables. Furthermore, the paths generated by the estimated HMM facilitate the detection of change points which are interpreted as shifts in the observed distribution. This capability is an innovative feature of our framework, significantly enhancing its ability to identify structural changes within the data.

In the case of empirical data analysis, the Bayesian Information Criterion (BIC) fails to discriminate between scenarios characterized by different priors, suggesting that the model is largely insensitive to the choice of these priors.

This issue is well known in the literature and reflects inherent limitations of the BIC, especially in settings involving Hidden Markov Models -- particularly when the latent states are highly overlapping — or, more generally, in mixture models, where model complexity may not be sufficiently penalized in the presence of overlapping components. More specifically, mixture models and HMMs often exhibit irregularities, such as local identifiability issues and singularities in the likelihood function, which can partially invalidate the theoretical assumptions underlying the derivation of the information criterion. In such cases, the adoption of alternative criteria (e.g., integrated complete likelihood, marginal likelihood via bridge sampling) can help disambiguate certain situations; however, these approaches do not offer a definitive solution.

With reference to the empirical application discussed in the previous section, we implemented the integrated complete likelihood, which on the one hand exhibited greater sensitivity to the prior specifications, and on the other provided results that did not consistently favor the two-state specification.

The results of the two- and three-state simulations underscore the strengths of the proposed framework with skew-normal components. In both scenarios, the model exhibits accurate parameter recovery, with consistent and reliable estimation of location, scale, and skewness parameters. This confirms the model’s ability to capture the asymmetric features inherent in the data. The transition matrices are also effectively estimated, allowing for precise detection of regime changes. In addition, the confusion matrices and the corresponding metrics (accuracy and Kappa) further validate the model's ability to accurately classify observations into hidden states.

In general, incorporating skew-normal distributions into the model enhances its flexibility to accommodate asymmetry in the data, offering a more realistic representation of complex temporal patterns. When combined with the HMM framework, this feature ensures the accurate identification of structural breaks and regime changes, even in the presence of heterogeneity and non-Gaussian behaviors. These findings demonstrate the suitability of the proposed approach for a wide range of applications involving temporal data characterized by asymmetric features and regime-switching dynamics. Given its capabilities, the approach is particularly well suited for studies that require a high level of flexibility in the target distribution. A further avenue for research could address the well-known challenge of selecting an appropriate number of states in HMMs, going beyond the Bayesian Information Criterion or Integrated Completed Likelihood criterion and emphasizing common pitfalls and practical challenges when analyzing real-world data. This issue remains a prominent topic in the literature, with solutions often varying significantly depending on the data set and the specific application domain.

We have also addressed the potential identifiability issues of the model, which become particularly delicate when the mixture is extended beyond the Gaussian case to include components with asymmetry (skewness). The problem raised by \cite{Teicher63} has been clarified, specifically in the context of skew-normal distributions, by the more recent contributions of \cite{WZW23}. However, this topic still deserves further investigation in future work.

\section*{Acknowledgment}

The authors are grateful for insightful suggestions from two anonymous reviewers. 

\section*{Statements and Declarations}

The authors declare that they have no financial or non-financial conflicts of interest. During the preparation of this work, the authors used WRITEFULL in overleaf to improve the English wording. After using this tool, the authors reviewed and edited the content as needed and take full responsibility for the content of the publication.

\bibliographystyle{agsm}
\bibliography{HMM_SkN}

\end{document}